\documentclass[preprint]{aastex63}
\usepackage[utf8]{inputenc}
\usepackage[T1]{fontenc}
\usepackage{palatino}
\usepackage{times,amssymb}
\usepackage{amsmath}
\usepackage{units}
\usepackage{graphicx}
\usepackage{wrapfig}
\usepackage{natbib}
\usepackage[nottoc]{tocbibind}
\usepackage[titletoc,toc,title]{appendix}
\usepackage{bm}

\begin{document}

\title{Thermal and orbital evolution of low-mass exoplanets}

\correspondingauthor{Michaela Walterov\'{a}}
\email{kanova@karel.troja.mff.cuni.cz}

\author[0000-0002-6779-3848]{Michaela Walterov\'{a}}
\affiliation{Charles University, Faculty of Mathematics and Physics, Department of Geophysics \\
V Holešovičkách 2\\
180 00 Prague, Czech Republic}

\author[0000-0001-8227-0685]{Marie B\v{e}hounkov\'{a}}
\affiliation{Charles University, Faculty of Mathematics and Physics, Department of Geophysics \\
V Holešovičkách 2\\
180 00 Prague, Czech Republic}

\received{May 31, 2020}
\revised{July 17, 2020}
\accepted{July 21, 2020}
\submitjournal{\apj}

\begin{abstract}

Thermal, orbital, and rotational dynamics of tidally loaded exoplanets are interconnected by intricate feedback. The rheological structure of the planet determines its susceptibility to tidal deformation and, as a consequence, participates in shaping its orbit. The orbital parameters and the spin state, conversely, control the rate of tidal dissipation and may lead to substantial changes of the interior. We investigate the coupled thermal-orbital evolution of differentiated rocky exoplanets governed by the Andrade viscoelastic rheology. The coupled evolution is treated by a semi-analytical model, 1d parametrized heat transfer and self-consistently calculated tidal dissipation. First, we conduct several parametric studies, exploring the effect of the rheological properties, the planet’s size, and the orbital eccentricity on the tidal locking and dissipation. These tests show that the role of tidal locking into high spin-orbit resonances is most prominent on low eccentric orbits, where it results in substantially higher tidal heating than the synchronous rotation. Second, we calculate the long-term evolution of three currently known low-mass exoplanets with nonzero orbital eccentricity and absent or yet unknown eccentricity forcing (namely GJ 625 b, GJ 411 b, and Proxima Centauri b). The tidal model incorporates the formation of a stable magma ocean and a consistently evolving spin rate. We find that the thermal state is strongly affected by the evolution of eccentricity and spin state and proceeds as a sequence of thermal equilibria. Final despinning into synchronous rotation slows down the orbital evolution and helps to maintain long-term stable orbital eccentricity.
\end{abstract}


\section{Introduction} \label{sec:intro}

Internal dynamics of close-in exoplanets and large moons in the Solar System are closely linked to their tidal interaction with the primary. As an effective heat source, tidal dissipation can sustain liquid oceans under the surface of large icy moons \citep[e.g.,][]{hussmann10,chen14} or maintain the extreme volcanic activity of Jupiter's moon Io \citep[e.g.,][]{peale79,segatz88}. Beyond the realms of the Solar System, tidal heating is believed to transform close-in rocky exoplanets into lava worlds \citep[e.g.,][]{behounkova11,barr18,henning18}, influence their tectonic regime \citep{zanazzi19}, alter the boundaries of the conventional habitable zone for exoplanets or exomoons \citep[e.g.,][]{jackson08,heller13,dobos17,renaud18}, and it is also one of the suggested mechanisms responsible for the existence of inflated hot Jupiters \citep[e.g.,][]{bodenheimer2001,jermyn17}.

Tidal loading and the subsequent transfer of angular momentum and orbital energy is also the cause of long-term orbital evolution, accompanied by the disturbed body's despinning into spin-orbit synchronization, pseudosynchronization or a higher spin-orbit resonance \citep[e.g.,][]{ferrazmello13,makarov13,correia14,ferrazmello15}. As a consequence of the spin-orbital dynamics, close-in moons and exoplanets are expected to tend towards circular orbit and synchronous rotation, unless they are disturbed by other bodies in the system or by the tidal response of a rapidly rotating primary \citep[e.g.,][]{bolmont16}. The rate of tidally induced orbital evolution depends on the moon’s or planet’s ability to dissipate mechanical energy. Different interiors support different dissipation mechanisms. While the dissipation inside terrestrial bodies or rocky parts of gas giants is dominated by large-scale viscous flow \citep{ferrazmello13}, jovian worlds are typically heated by inertial waves and turbulent convection in their massive atmospheres \citep{ogilvie04}. Owing to their higher tidal quality factor $Q$, they are also a few orders of magnitude less susceptible to tidally-induced spin and orbital evolution \citep{goldsot66}. Since the interior structure and dissipation mechanisms of extremely heated bodies depend on the varying interior temperature \citep[e.g.,][]{henning09,renaud18}, the rate of orbital evolution is presumably also a function of time.\\

The feedback between the thermal and orbital evolution has been investigated particularly in the context of large Solar System satellites. \citet{ojakangas86} assessed mutual interconnection between the varying interior temperature of Jupiter’s moon Io and the evolution of its orbital eccentricity, which is affected both by tides and by the Laplace resonance with other Galilean satellites. Assuming heat loss by mantle convection and heat generation by viscoelastic tidal dissipation, they described a process of periodic cooling down and heating up of the satellite, controlled by inverse dependence of the tidal quality function $k/Q$ on the temperature. Several years later, \citet{fischer90} extended the model by considering partial melting of the interior. The presence of melt decreases the moon’s average rigidity and further reduces the tidal dissipation. In addition to the oscillatory state described by \citet{ojakangas86}, the authors identified an approximate thermal and dynamical equilibrium, in which the moon can be temporarily stabilized. The equilibrium is characterized by a very low rate of change in the eccentricity and interior temperature. Further studies of coupled thermal-orbital evolution with more complex models of the interior were since then presented by a number of authors \citep[e.g.,][]{hussmann04,tobie05b,neveu19}.

Partial melting is likely to be an important regulating mechanism also in the case of close-in terrestrial exoplanets. Identically to the case of Solar System satellites, the emergence of melt yields an abrupt change of rheological parameters and decreased tidal dissipation, which prevents runaway heating of the mantle \citep{makarov18}. For multiplanetary systems in mean motion resonances, the coupled thermal-orbital evolution follows a pattern similar to the evolution of large Saturnian or Jovian satellites. In single-planetary systems or systems without substantial eccentricity forcing, on the other hand, the decreased dissipation may be reflected in the unexpected orbital parameters. Specifically, it has been argued \citep[e.g.,][]{henning09,henning14,makarov15} that partial melting may explain nonzero orbital eccentricities of exoplanets for which the standard tidal theories predict rapid circularization.

\citet{shoji14} and \citet{driscoll15} investigated the long-term thermal-orbital evolution of single-planetary systems around low-mass stars. Both studies focused on small, synchronously rotating exoplanets in the habitable zone and implemented a combined model of parametrized mantle convection with global melting, viscoelastic tidal dissipation and simplified orbital evolution. \citet{shoji14} assumed stagnant lid convection regime and predicted two possible evolution branches of the planets in question. Depending on the initial orbital eccentricity and the stellar mass, the planet either undergoes runaway cooling, with increasing mantle viscosity and gradually terminating mantle convection, or it is affected by runaway heating, which is eventually stopped by partial melting. In either of these states, the semi-major axis and the orbital eccentricity change very slowly over several billions of years and may allow the planet to remain habitable for a considerable time. In contrast, \citet{driscoll15} prescribed mobile lid regime, in which the planet's lithosphere participates in the convection. As a consequence of more efficient energy dissipation, they found rapid decay of the eccentricity of close-in exoplanets and relatively low importance of tidal heating on long timescales. The authors also performed plenty of parametric studies, in which they illustrated the complex dependence of tidal heat rate and other quantities on the initial orbital parameters.

While the assumption of synchronous rotation, taken in the cited studies, is well justified in low eccentricity cases, the evolution of exoplanets on eccentric orbits is most probably marked by tidal locking into higher than synchronous resonances or pseudosynchronous rotation \citep[e.g,][]{dobro07,ferrazmello13,correia14}. A self-consistent model of an eccentric exoplanetary system requires not only the coupling between thermal and orbital evolution but also the simultaneous assessment of the stable spin state. In the specific case of the multiplanetary system TRAPPIST-1, the influence of a tidally induced change of interior properties on the stability of higher spin states has been discussed by \citet{makarov18}. The authors identified that, as a consequence of an abrupt change in the mantle’s rheological parameters, the planet might leave its initially high spin-orbit resonance and evolve towards synchronization or towards pseudosynchronization, expected for molten rocky worlds on eccentric orbits \citep{makarov15}. Both the melting and the change of equilibrium spin state result in substantially decreased tidal dissipation.\\

The purpose of this paper is twofold. First, we aim at providing a parametric study of tidal locking and tidal heating for close-in terrestrial exoplanets governed by Andrade rheology. Second, we investigate the coupled thermal and orbital evolution of model bodies inspired by three currently known low-mass exoplanets with nonzero eccentricity. The former goal is achieved by numerically solving the tidal despinning and dissipation for planets with constant rheological parameters. The latter goal is accomplished by implementing a semi-analytical model including spin-orbital evolution of a layered viscoelastic planet with emerging subsurface magma ocean, self-consistently calculated tidal heat generation and simplified, parametrized mantle convection in stagnant-lid regime. We begin by introducing the evolution equations used for the calculation of time-dependent orbital elements and spin rate (Section \ref{sec:orbit}). Then, in Section \ref{sec:rheo}, we briefly comment the chosen viscoelastic rheology and outline the calculation of tidal heating. Section \ref{sec:thermal} is dedicated to the thermal evolution of rocky exoplanets, namely the parametrized stagnant lid convection with partial melting. Section \ref{sec:num} gives an overview of the composite model's numerical implementation.

In the subsequent three sections, we present the results of the study with either constant or evolving rheological parameters. Section \ref{sec:results1} illustrates the effect of various model parameters on the tidal heating and the highest stable spin state of a generic Earth-mass exoplanet. After discussing the main features of the parametric dependencies, we conduct the same kind of parametric study for exoplanets Proxima Centauri b, GJ 625 b, and GJ 411 b (Section \ref{sec:results2}). The outcome of the coupled thermal-orbital model for the three exoplanets is presented in Section \ref{sec:results3}. We discuss the obtained results in Section \ref{sec:disc} and conclude with Section \ref{sec:concl}. The paper also contains two appendices. In Appendix \ref{app1}, we give a brief description of the normal mode theory, which is used for the computation of tidal deformation in a layered body, and Appendix \ref{app2} helps to explain the parameter dependence of tidal torque.

\section{Orbital evolution} \label{sec:orbit}

Our model system consists of a spherical star with mass $m_*$ and a single rocky planet with mass $m_{\rm{p}}\ll m_*$, whose trajectory is defined by the semi-major axis $a$ and the eccentricity $e$. The planet, considered here as an extended body, deforms in the heterogeneous gravitational field of the host star and its orbital parameters undergo secular tidal evolution. Additional potential due to the planet's deformation, which presents a perturbation to the standard two-body problem \citep[e.g.][]{murray99}, can be expressed in the form of a disturbing function. Following the linear tidal theory developed by \citet{darwin80} and \citet{kaula61}, we expand the disturbing function into a Fourier serie in space and time and insert it into Lagrange planetary equations for the semi-major axis and the eccentricity. The secular evolution equations are then written as a linear combination of individual tidal modes \citep{kaula64}:

\footnotesize
\begin{equation}
\dot{a} = -\sum_{lmpq} \frac{2 \mathcal{G} m_*}{na}\; \frac{R^{2l+1}}{a^{2l+2}}\; (l-2p+q)\; \frac{(l-m)!}{(l+m)!}\; (2-\delta_{0m})\; \big[\mathcal{G}_{lpq}(e)\big]^2\; \big[\mathcal{F}_{lmp}(\beta)\big]^2\; k_l\; \sin\varepsilon_{lmpq}\; , \label{eq:kaula_a}
\end{equation}
\begin{equation}
\dot{e} = -\sum_{lmpq} \frac{\mathcal{G} m_*}{na^2}\; \frac{\sqrt{1-e^2}}{e}\; \frac{R^{2l+1}}{a^{2l+2}}\; \Big[\sqrt{1-e^2}(l-2p+q)-(l-2p)\Big]\; \frac{(l-m)!}{(l+m)!}\; (2-\delta_{0m})\; \big[\mathcal{G}_{lpq}(e)\big]^2\; \big[\mathcal{F}_{lmp}(\beta)\big]^2\; k_l\; \sin\varepsilon_{lmpq}\; . \label{eq:kaula_e}
\end{equation}\normalsize\\
In the above equations, $\mathcal{G}$ is the Newton's gravitational constant, $n=\sqrt{\mathcal{G}(m_*+m_{\rm{p}})/a^3}$ stands for the planet's mean motion and $R$ symbolizes the planet's outer radius. Quantities $\mathcal{G}_{lpq}(e)$ and $\mathcal{F}_{lmp}(\beta)$ are Kaula's functions of orbital eccentricity and inclination relative to the planet's equator \citep[e.g.,][]{kaula61,kaula64,allan65}. In our case, the "inclination" is equal to the planet's obliquity $\beta$, which we set to zero. The tidal response of the planet, determined by its rheological properties and interior structure, is represented by the frequency dependent tidal Love number $k_l = k_l(\omega_{lmpq})$ and by the tidal phase lag $\varepsilon_{lmpq} = \varepsilon_{lmpq}(\omega_{lmpq})$. While the former quantifies the ratio between the amplitude of the additional potential and the amplitude of the tidal potential, the latter characterizes the lagging between the two potentials in the frequency domain. When working with viscoelastic models of the interior, it is also appropriate to introduce the complex Love number $\bar{k_l}(\omega_{lmpq})$ \citep[e.g.,][]{castillo11}, whose relation to the two quantities is

\footnotesize
\begin{equation}
\bar{k_l}(\omega_{lmpq}) = k_l\; \exp\big\{-i\varepsilon_{lmpq}\big\}\; .
\end{equation}\normalsize\\
Finally, the frequencies of the individual modes $\{l,m,p,q\}$ are

\footnotesize
\begin{equation}
\omega_{lmpq} = (l-2p+q) n - m \dot{\theta}\; , \label{eg:freqs}
\end{equation}\normalsize\\
where $\dot{\theta}$ stands for the planet's spin rate. Similarly to the calculation of the semi-major axis and the eccentricity, the secular evolution of the spin rate can be also written as a sum of individual modes \citep[e.g.,][]{dobro07,efroimsky09},

\footnotesize
\begin{equation}
\ddot{\theta} = \frac{\mathcal{G} m_*^2}{C}\; \sum_{lmpq}  \frac{R_{\rm{p}}^{2l+1}}{a^{2l+2}}\; \frac{(l-m)!}{(l+m)!}\; (2-\delta_{0m})\; m\; \big[\mathcal{G}_{lpq}(e)\big]^2\; \big[\mathcal{F}_{lmp}(\beta)\big]^2\; k_l\; \sin\varepsilon_{lmpq}\; , \label{eq:rot}
\end{equation}\normalsize\\
with $C$ being the principal moment of inertia with respect to the rotational axis. For the sake of simplicity, we set $C$ equal to the moment of inertia of sphere with the same mass and radius as the planet. While it is possible to accommodate the Darwin-Kaula theory also for the study of secular evolution of the planet's obliquity \citep{boue19}, we do not include this effect in our model and consider only the planar case with $\beta=0\degr$. Nonzero initial obliquity would temporarily affect the stability of higher than synchronous spin-orbit resonances \citep{boue16} and would be an additional source of tidal heating, complementing the eccentricity tides \citep[e.g.,][]{peale78}. However, when the planet's spin-orbital evolution is shaped only by tides, the obliquity in stable spin states usually tends towards zero \citep{boue16}.

For the sake of completeness, we note that the spin and orbital evolution might be induced also by the deformation of the star under the gravitational action of the planet. However, due to the large difference between masses and to lower typical dissipation rates in stars, compared to the terrestrial planets \citep[e.g.,][]{hansen10}, the star's contribution to the system's tidal evolution is neglected in this study and we consider only the dissipation in the companion. Nevertheless, tides raised by the planet on the host star play an important role in the evolution of hot Jupiters orbiting fast rotating stars \citep{bolmont16}, in which case they should be taken into account.

\section{Tidal deformation} \label{sec:rheo}

The reaction of a tidally loaded exoplanet with a given mass and radius is determined by its interior structure and rheological properties. The mineralogical composition of the mantle, as well as the existence of a liquid core or a subsurface ocean, translates into the previously introduced complex Love numbers $\bar{k_l}(\omega_{lmpq})$. For a homogeneous body with averaged interior properties, the complex Love numbers can be expressed by a relatively simple analytical formula \citep{castillo11}. This approach facilitates the qualitative examination of the problem; however, the assumption of a homogeneous interior might not always be justified. As has been shown by a number of studies \citep[e.g.,][]{castillo11,henning14,folonier15,tobie19}, the tidal deformation and dissipation vary between different models of interior structure and the radial stratification cannot be generally neglected.

Here, we focus on planets with liquid core and emerging magma ocean; i.e., we are concerned with a layered interior. To calculate the complex Love numbers of a differentiated planet, we follow in the steps of the previous tidal studies and adopt the normal mode theory \citep[e.g.,][]{takeuchi72,wu82,sabadini04,tobie05}. A draft of the method is presented in Appendix \ref{app1}. This calculation assumes that each interior layer is endowed with its own material and rheological properties. Specifically, the mantle is described by linear viscoelastic rheology, which predicts instantaneous deformation on seismological timescales and gradual creeping on geological timescales. In the following, we discuss the chosen rheological model and the calculation of tidal dissipation.

\subsection{Rheological models}

According to the principle of correspondence, the equations of motion for a linear viscoelastic continuum in the frequency domain are analogous to the equations governing the motion of an elastic material. The static rigidity $\mu$ only needs to be rewritten to its complex and frequency-dependent counterpart $\bar{\mu}(\omega)$. Similarly to the elastic case, the complex rigidity characterizes the relation between the deviatoric part of the incremental strain tensor $\bar{\boldsymbol{\varepsilon}} = \frac12 \big(\nabla\mathbf{u}+\nabla^{t}\mathbf{u}\big)$ and the deviatoric part of the incremental Cauchy stress tensor $\bar{\boldsymbol{\sigma}}$,

\footnotesize
\begin{equation}
\bar{\boldsymbol{\sigma}} = 2\bar{\mu}(\omega)\, \bar{\boldsymbol{\varepsilon}}\; ,
\end{equation}\normalsize\\
where $\mathbf{u}$ stands for the incremental displacement. Alternatively, we may introduce complex compliance $\bar{J}(\omega)$, defining the will of a material to deform under applied stress,

\footnotesize
\begin{equation}
2\bar{\boldsymbol{\varepsilon}} = \bar{J}(\omega)\, \bar{\boldsymbol{\sigma}}\; .
\end{equation}\normalsize\\
Depending on the underlying mechanisms of deformation, the material's response can be described by one of many rheological models. The simplest viscoelastic model used in planetary science \citep[e.g.,][]{ferrazmello13,correia14} is the Maxwell model, with complex compliance given by

\footnotesize
\begin{equation}
\bar{J}(\omega) = \frac{1}{\mu} - \frac{i}{\eta\omega}\; . \label{eq:maxwell}
\end{equation}\normalsize\\
While the first term in equation (\ref{eq:maxwell}) accounts for the instantaneous, elastic deformation of the material, the second term, depending on the frequency and on the viscosity $\eta$, stands for gradual viscous creep. The Maxwell rheology is well suited for the description of geophysical phenomena acting on long timescales, such as global isostatic adjustment or Chandler wobble. When applied to tides, it performs reasonably well at low frequencies. However, at high tidal frequencies, the Maxwell model tends to underestimate the attenuation in the medium and, consequently, the tidal dissipation \citep{efroimsky07}. 

Owing to the variety of deformation mechanisms observed in real solids, the accurate description of the planet's response requires the introduction of more complex rheological models, consistent with laboratory experiments and seismological or geodetical measurements \citep[for an overview, see, e.g.,][]{efroimsky07,henning09,castillo11}. The best fit to experimental data for polycrystalline material is presented by three models: Andrade \citep{andrade10}, extended Burgers \citep{faul05}, and Sundberg-Cooper \citep{sundberg10}, each of which entails different anelastic extension to the simple Maxwell-like viscoelastic behaviour. The desire to keep the number of model parameters at minimum while retaining a sufficiently accurate description of the deformation leads us to prefer the Andrade rheology, whose complex compliance is

\footnotesize
\begin{equation}
\bar{J}(\omega) = \frac{1}{\mu} - \frac{i}{\eta\omega} + \frac{\mu^{\alpha-1}}{(i\,\zeta\eta\omega)^{\alpha}}\,\Gamma(1+\alpha)\; . \label{eq:andrade}
\end{equation}\normalsize\\
The last term in equation (\ref{eq:andrade}) stands for a transient, anelastic creep, which dominates the material's response at high\footnote{According to \citet{karato90}, Andrade rheology is applicable to the Earth's response at frequencies higher than $\sim\unit[1]{yr^{-1}}$. However, the exact position of the frequency threshold between the anelastic and viscoelastic regimes depends exponentially on the temperature and may vary greatly with the thermal state of the mantle.} frequencies. Symbols $\alpha$ and $\zeta$ stand for empirical parameters characterizing the duration of transient creep and the ratio of material's relaxation time to the Maxwell time ${\eta}/{\mu}$, respectively. Both parameters depend on the prevalent deformation mechanism at given stresses, temperatures and chemical compositions \citep{karato90}.\\

A characteristic feature of viscoelastic tidal models is the occurrence of distinct stable spin states, i.e. distinct stationary solutions to equation (\ref{eq:rot}), associated either with spin-orbit resonances or with pseudosynchronous rotation \citep[e.g.,][]{correia14,ferrazmello15,makarov15}. Additionally, complex viscoelastic models are endowed with increased tidal heating at high frequencies and they enable the planet to remain tidally active for long periods \citep{renaud18}.

\subsection{Tidal heating} \label{sec:thermal_tidal}

Periodic deformation of a viscoelastic body is accompanied by the dissipation of mechanical energy, which results in tidal heating. The average heat rate produced by the dissipation in the entire volume of the planet over one orbital period can be written as \citep{segatz88,efroimsky14}

\footnotesize
\begin{equation}
\bar{P}^{{\rm{tide}}} = -\sum_{l=2}^{\infty}\frac{(2l+1)\,n}{8\pi^2\mathcal{G}R}\;\int_0^{T_{\rm{orb}}}\int_S \delta\Phi_l(R,\vartheta,\varphi,t) \;\frac{\partial\Phi_l(R,\vartheta,\varphi,t)}{\partial t}\, {\rm{d}}S\, {\rm{d}}t \, , \label{eq:segatz}
\end{equation}\normalsize\\
where $\Phi_l$ and $\delta\Phi_l= |\bar{k}_l(\omega)|\,\Phi_{l,\rm{lag}}$ are the degree-$l$ tidal and disturbing potentials evaluated at the planet's surface. The subscript "lag" indicates that the argument of the disturbing potential should be complemented with the previously introduced tidal phase lag. Expressing the two potentials in the form of a Darwin-Kaula expansion \citep{kaula61,kaula64} and making use of the orthogonality of associated Legendre polynomials, we can rewrite the global tidal heat rate (\ref{eq:segatz}) into the analytical form \citep{efroimsky14}

\footnotesize
\begin{equation}
\bar{P}^{{\rm{tide}}} = -\frac{\mathcal{G}m_{*}^2}{a} \sum_{lmpq}\left(\frac{R}{a}\right)^{2l+1}\; (2-\delta_{m0})\, \frac{(l-m)!}{(l+m)!}\,\big[\mathcal{G}_{lpq}(e)\big]^2\; \big[\mathcal{F}_{lmp}(\beta)\big]^2\; \omega_{lmpq}\;{\rm{Im}}\big\{\bar{k}_l(\omega_{lmpq})\big\}\; . \label{eq:segatz_final}
\end{equation}\normalsize\\

Note that equation (\ref{eq:segatz_final}) holds for an arbitrary obliquity and orbital eccentricity as well as for arbitrary spin rate, provided that the overall deformation can be described by a linear tidal theory.

\section{Thermal evolution} \label{sec:thermal}

The thermal state of planetary bodies is controlled by a combination of heating and cooling mechanisms, differing in their significance and in their characteristic timescales. Internal heat sources include remnant gravitational energy released at the time of the planet's formation and differentiation, latent heat extracted during phase transitions, radiogenic heating of the crust and mantle, and tidal dissipation. Secular cooling of the planetary interior is realized mainly by mantle convection and conduction, depending on the size, temperature gradient, and rheological properties of the mantle.\\

The following section contains several important assumptions. Keeping in mind the wealth of possible thermal histories of the exoplanets, including, for instance, the occurrence of plate tectonics or episodic resurfacing events, we focus specifically on the stagnant lid convection. This gives us the advantage of a relatively simple parametric description, allowing for a systematic parametric study. Furthermore, given the lack of information on the tectonic regimes of exoplanets and the scarcity of plate tectonics in the Solar System, the stagnant lid convection is often considered as a conservative guess \citep[e.g.,][]{shoji14,tosi17}.

Since the main subject of this work is the evolution of strongly tidally loaded exoplanets, we also restrict the mantle heating mechanisms to the initial core-mantle temperature difference and to the volumetric tidal dissipation. The contribution of latent heat to the overall energy balance is, however, included in the adopted equations. Although the radiogenic heating may be an important source in the elastic lithosphere (and crust) and it might slow down the cooling of the mantle, its contribution would be most pronounced in the initial stages of the evolution, which are, however, also dominated by tidal heating.

The last important assumption is the absence of melt extraction from the mantle. Low efficiency of the heat transport through the stagnant lid, together with immense heating of planets on eccentric orbits, may lead to partial melting of the interior and formation of a subsurface magma ocean. Current studies using parametrized stagnant lid convection consider either perfect mixing of the interior \citep[e.g.,][]{henning09,shoji14,driscoll15} or instantaneous melt extraction accompanied by crustal production \citep[e.g.,][]{breuer06,tosi17}. Although the subsurface magma can be extracted from the interior by heat pipe volcanism \citep{spohn91,moore17}, we do not consider any melt transport mechanisms in this study, and we instead include the effect of the emerging magma ocean into the tidal model.

\subsection{Parametrized mantle convection}

To inspect the long-term thermal evolution of the planet, we adopt a 1d parametrized model of mantle convection in the stagnant lid regime \citep[e.g.,][]{breuer06,grott08}. The evolution of the temperature on the top of the convecting mantle $T_{{\rm{m}}}$ and at the core-mantle boundary $T_{{\rm{c}}}$ is governed by the energy balance in the tidally heated planet \citep{breuer06},

\footnotesize
\begin{equation}
\rho_{{\rm{m}}} c_{{\rm{m}}} V_{{\rm{m}}} (1+{St}) \frac{{\rm{d}}T_{{\rm{m}}}}{{\rm{d}}t} = -q_{{\rm{m}}}A_{{\rm{m}}} + q_{{\rm{c}}} A_{{\rm{c}}} + \bar{P}^{{\rm{tide}}}\; , \label{eq:tempeqM}
\end{equation}

\begin{equation}
\rho_{{\rm{c}}} c_{{\rm{c}}} V_{{\rm{c}}} \frac{{\rm{d}}T_{{\rm{c}}}}{{\rm{d}}t} = - q_{{\rm{c}}} A_{{\rm{c}}}\; , \label{eq:tempeqC}
\end{equation}\normalsize\\
where $\rho_{{\rm{m}}}$ and $\rho_{{\rm{c}}}$ is the mean density in the mantle and in the core, respectively, and $c_{{\rm{m}}}$ and $c_{{\rm{c}}}$ are the corresponding specific heat capacities. Symbol ${St}$ stands for the Stefan number related to the latent heat $L_{\rm{m}}$ consumed or generated during partial melting or solidification,

\footnotesize
\begin{equation}
{St} = \frac{L_{\rm{m}}}{c_{{\rm{m}}}}\, \frac{{\rm{d}}\phi_{\rm{m}}}{{\rm{d}}T_{\rm{m}}}\; , \label{eq:stefan}
\end{equation}\normalsize\\
where $\phi_{\rm{m}}$ signifies the total melt fraction in the mantle. Additionally, $A_{{\rm{c}}}$, $A_{{\rm{m}}}$, $V_{{\rm{c}}}$ and $V_{{\rm{m}}}$ are the total surface areas and volumes of the core and the mantle, and $q_{{\rm{c}}}$ and $q_{{\rm{m}}}$ are the heat fluxes from the core to the mantle and from the mantle to the lithosphere, respectively. The last two quantities can be expressed as

\footnotesize
\begin{equation}
q_{{\rm{c}}} = k_{{\rm{m}}}\; \frac{T_{{\rm{c}}}-T_{{\rm{b}}}}{\delta_{{\rm{c}}}} \label{eq:qc}
\end{equation}\normalsize
and
\footnotesize
\begin{equation}
q_{{\rm{m}}} = k_{{\rm{m}}}\; \frac{T_{{\rm{m}}}-T_{{\rm{l}}}}{\delta_{{\rm{u}}}}\; , \label{eq:qm}
\end{equation}\normalsize\\
where $k_{{\rm{m}}}$ is the thermal conductivity of the mantle, $T_{{\rm{b}}}$ is the temperature on the bottom of the convecting mantle, $T_{{\rm{l}}}$ stands for the temperature at the base of the lithosphere and $\delta_{{\rm{c}}}$ and $\delta_{{\rm{u}}}$ are the thicknesses of the lower and upper thermal boundary layers, given by the boundary layer theory. The temperatures throughout the convective mantle follow the adiabatic profile. According to the chosen stagnant lid parametrization, the boundary layer thicknesses should be determined by the ratio of the local Rayleigh number to the critical Rayleigh number \citep{tosi17}. However, as explained in the following section, we calculate the thickness of the upper boundary layer using the "mean Rayleigh number", corresponding to the average mantle viscosity. This choice enables us to mimic the role of a magma ocean.\\

In the presented model, the thermal evolution of the planet affects the interior structure and the tidal response in three ways. First, the increasing or decreasing temperature at the top of the mantle regulates the heat flux into the lithosphere and, as a result, determines the thickness of the stagnant lid $D_{\rm{l}}$ \citep{grott08},

\footnotesize
\begin{equation}
\rho_{{\rm{m}}} c_{{\rm{m}}} (T_{\rm{m}}-T_{\rm{l}}) \frac{{\rm{d}}D_{{\rm{l}}}}{{\rm{d}}t} = -q_{{\rm{m}}} - k_{{\rm{m}}} \frac{\partial T}{\partial r}\bigg\rvert_{r=R_{\rm{l}}}\; . \label{eq:del_lid}
\end{equation}\normalsize\\
The second term on the right-hand side, which is being evaluated at the stagnant lid base with radius $R_{\rm{l}}$, can be obtained analytically \citep[e.g.][]{carslaw59} from the heat conduction equation

\footnotesize
\begin{equation}
\frac{1}{r^2} \frac{\partial}{\partial r} \left(r^2 k_{\rm{l}} \frac{\partial T}{\partial r}\right) = 0\; ,
\end{equation}\normalsize\\
where we have neglected heat sources in the elastic---and thus non-dissipative---lithosphere. The surface temperature $T_{\rm{s}}$ of the model planets is held constant.

As the second of the coupling mechanisms, the variations of the interior temperature influence the rheological properties of mantle minerals. Specifically, the temperature dependence of the local mantle viscosity $\eta$ can be expressed by the Arrhenius law as

\footnotesize
\begin{equation}
\eta(T) = \eta_{0} \exp\left(\frac{A}{R_{\rm{gas}}}\frac{T_0-T}{T_0 T}\right)\; , \label{eq:arreta}
\end{equation}\normalsize\\
where $\eta_0$ is the reference viscosity at reference temperature $T_0=\unit[1600]{K}$ \citep{grott08}, $A$ stands for the activation energy and $R_{\rm{gas}}$ is the gas constant. In addition to the temperature dependence, mantle rheology should be also determined by the local pressure, whose role in shallow depths of the mantle is to increase the viscosity. The effect of extreme pressures in the lower mantle of massive terrestrial exoplanets is, however, still a question of debate \citep[e.g.,][]{karato11,stamenkovic12}. Here, we do not include the pressure dependence of the mantle viscosity explicitly, but we rather assume several different values of the reference viscosity $\eta_0$ in order to cover all possible viscosity models.

The third and last of the discussed mechanisms is partial melting in the shallow regions of the mantle, possibly followed by the formation of a magma ocean. The presence of melt, associated with an additional change in the local viscosity and rigidity, alters the planet's response to external loading and affects the efficiency of energy dissipation. The melt fraction at a given radius $\phi(r)$ depends on the local temperature $T(r)$, given by the mantle temperature profile, and on the chemical and mineralogical composition, which determines the solidus and liquidus temperatures throughout the planet,

\footnotesize
\begin{equation}
\phi(r) = \frac{T(r)-T_{\rm{sol}}(r)}{T_{\rm{liq}}(r)-T_{\rm{sol}}(r)}\; . \label{eq:meltfr}
\end{equation}\normalsize\\
The melting curves of mantle materials at pressures relevant to planetary science are determined by fitting laboratory data. Here, we follow \citet{monteux16}, who use empirical data from two experimental studies performed in two different pressure ranges and adjust the parameters of the fitted function to avoid discontinuities at the boundary of the two ranges. At pressures lower than $P=\unit[20]{GPa}$, the solidus and liquidus temperature can be obtained as \citep{herzberg96}

\footnotesize
\begin{equation}
T_{\rm{sol}} = 1661.2 \left(\frac{P}{1336\times 10^9}+1\right)^{{1}/{7.437}} \; ,
\end{equation}
\begin{equation}
T_{\rm{liq}} = 1982.1 \left(\frac{P}{6.594\times 10^9}+1\right)^{{1}/{5.374}}\; ,
\end{equation}\normalsize\\
while at pressures above $P=\unit[20]{GPa}$, deeper in the mantle, the following relations are used \citep{andrault11}:

\footnotesize
\begin{equation}
T_{\rm{sol}} = 2081.8 \left(\frac{P}{101.69\times 10^9}+1\right)^{{1}/{1.226}} \; ,
\end{equation}
\begin{equation}
T_{\rm{liq}} = 2006.8 \left(\frac{P}{34.65\times 10^9}+1\right)^{{1}/{1.844}} \; .
\end{equation}\normalsize\\
Once the local temperature exceeds the local solidus, the mantle rocks begin to melt. The molten material, first encapsulated in isolated cavities, gradually builds up a system of interconnected channels and as the local melt fraction reaches the disaggregation point $\phi_{\rm{D}}$ \citep[$40\%$-$60\%$; e.g.,][]{moore03}, it assumes the leading role in the rock's rheology. While the gradual formation of drops of partial melt in solid material does not substantially affect its rigidity and only accelerates the Arrhenius-like decrease in viscosity, trespassing of the disaggregation point is accompanied by a several orders of magnitude drop in both quantities. In order to characterize the described behaviour by a smooth and qualitatively adequate function, we adopt following dependence of the rheological parameters on the melt fraction

\footnotesize
\begin{align}
\log\mu(\phi) &= \log\mu_{\rm{max}} - \frac12 \left(\frac{2}{\pi}\arctan\frac{\phi-\phi_{\rm{D}}}{\Delta_{\mu}}+1\right)\, \log\frac{\mu_{\rm{max}}}{\mu_{\rm{min}}}\; , \label{eq:muphi}\\[2ex]
\log\eta(\phi,T) &= \log\eta(T) - \frac12 \left(\frac{2}{\pi}\arctan\frac{\phi-\phi_{\rm{D}}}{\Delta_{\eta}}+1\right)\, \log\frac{\eta(T)}{\eta_{\rm{min}}}\; , \label{eq:etaphi}
\end{align}\normalsize\\
where $\mu_{\rm{max}}$ and $\mu_{\rm{min}}$ is the rigidity of the solid and entirely molten rock, respectively, $\eta(T)$ is the temperature-dependent viscosity given by the Arrhenius law (\ref{eq:arreta}), $\eta_{\rm{min}}$ is the minimum viscosity, were it determined only by the melt fraction, and $\Delta_i$ stands for the disaggregation width of quantity $i$. Relations (\ref{eq:muphi}) and (\ref{eq:etaphi}) follow temperature and melt fraction dependence similar to the empirically justified expressions used in literature \citep[e.g.][]{fischer90,abe97,moore03}, while ensuring relatively steep but smooth parameter changes.

\section{Numerical implementation} \label{sec:num}

Long-term evolution of a planet in our model settings consists of processes with substantially different characteristic timescales. The shortest timescale is associated with the rotational evolution, and specifically with tidal despinning from the initial spin state to the closest equilibrium spin state (e.g., a spin-orbit resonance). Depending on the rate of tidal dissipation, the initial despinning operates on the scale of thousands or millions of years. On the other hand, the longest timescale is usually associated with the evolution of the semi-major axis and the orbital eccentricity. It may span from hundreds of millions to tens of billions of years. In between of these two extreme cases stands the mantle convection with a characteristic timescale of millions of years.

When choosing the adequate time step for the spin-orbital evolution model, it is necessary to take into account the precision of the calculation as well as its speed. The step should be short enough to correctly capture the changes of the spin rate and long enough to describe the long-scale processes in a reasonable time. In order to fulfil both of these requirements, we divide the calculation into two cycles. The \textit{short cycle} is dedicated to finding the equilibrium spin rate for a given semi-major axis, orbital eccentricity and interior structure, which are---on the short timescale---usually treated as constant. The \textit{long cycle}, on the other hand, takes steps in the orbital elements, assuming temporarily constant spin-orbit ratio.\\

\begin{figure}
  \begin{center}
    \includegraphics[width=1.\textwidth]{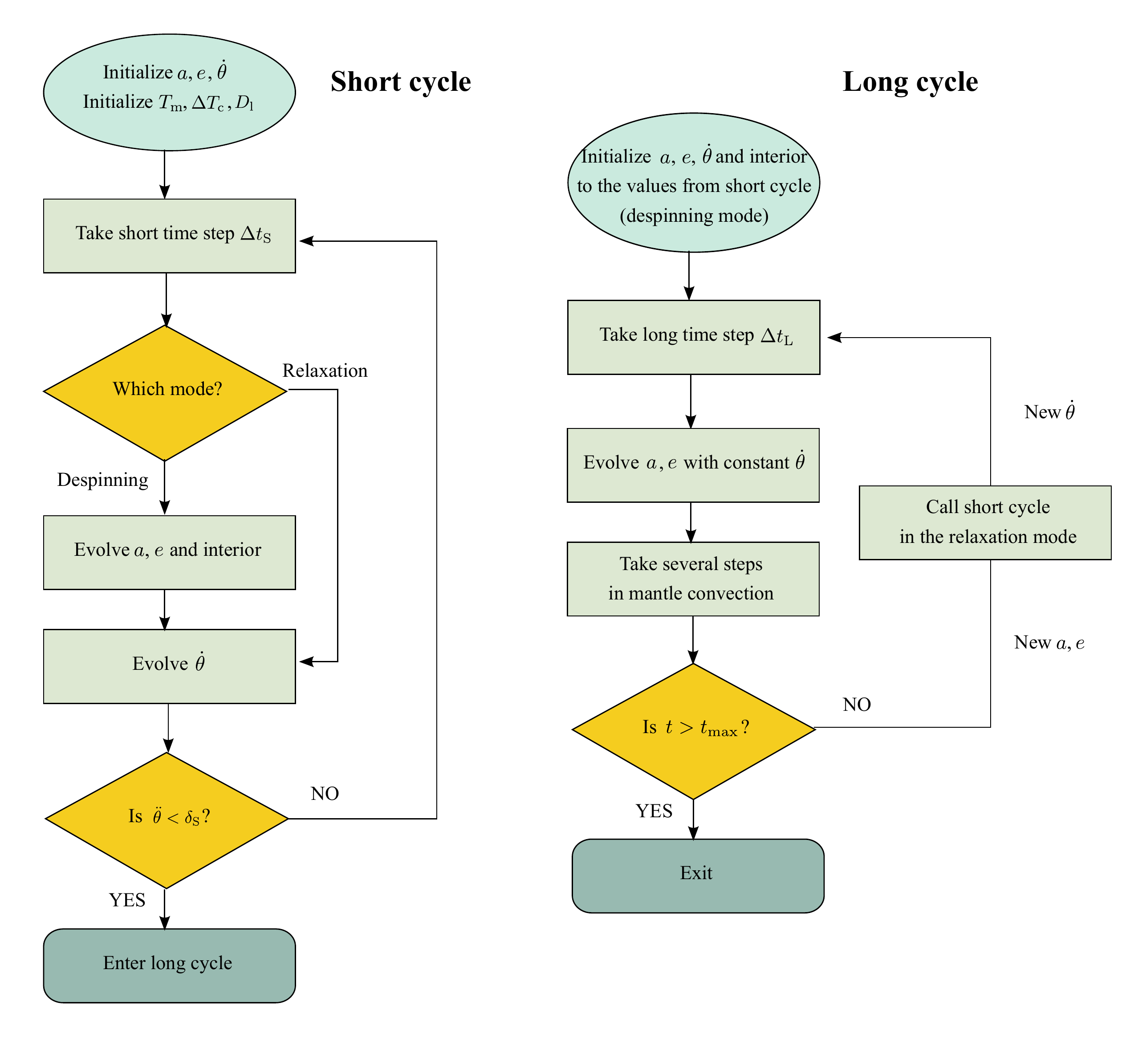}
  \end{center}
\caption{Schematic depiction of the computation flow. Short cycle proceeds with short time steps $\Delta t_{\rm{S}}$ and it is designed for finding the stable spin state. During its first run, it operates in the \textit{despinning mode}, which allows for the coupled evolution of the orbital elements and the spin rate. In later calls, it is set to the \textit{relaxation mode}, which serves to find the spin rate consistent with given (constant) semi-major axis and eccentricity. The long cycle takes long time steps $\Delta t_{\rm{L}}$ and calculates the evolution of the orbital parameters and thermal state for constant spin rate. During the computation, we alternate between steps in the long cycle and runs of the short cycle in the relaxation mode.} \label{fig:diagram}
\end{figure}

The flow of computation is schematically depicted in Figure \ref{fig:diagram} and proceeds as follows. In the beginning, we initialize the planetary and stellar masses, spin and orbital parameters and prescribe the planet's interior structure. The computation then starts with the short cycle in a specific \textit{despinning mode}. During the despinning, we evolve not only the spin rate but also the semi-major axis and the eccentricities, following equations (\ref{eq:kaula_a}) to (\ref{eq:rot}). Although the orbital parameters change very slowly on the short timescale, we include their evolution into the despinning mode in order to find the first stable spin state with high precision. Since the spin rate changes on relatively short timescales, the step size $\Delta t_{\rm{S}}$ in the short cycle is initialized to a few orbital periods. To solve the equations, we employ a 4th order predictor-corrector integration scheme (the Hamming's method) with variable step size controlled by the local error $\epsilon_{\rm{S}}$ in the spin rate $\dot\theta$ \citep[see, e.g.,][]{ralston65}.

The calculation in the short cycle is performed as long as the derivative $\ddot{\theta}$ exceeds a given limit $\delta_{\rm{S}}\times\dot{\theta}$. When the spin rate reaches the equilibrium value, we leave the short cycle and take one step in the long cycle. At this stage, the equilibrium spin-orbit ratio is considered constant and the step size $\Delta t_{\rm{L}}$ is set equal to the last step in the despinning mode of the short cycle. After the step in the long cycle, we recalculate the equilibrium spin rate using the short cycle in a \textit{relaxation mode}, in which the semi-major axis and the eccentricity are considered constant. For the rest of the computation, we alternate between taking one step in the long cycle and running the short cycle in the relaxation mode. Identically to the short cycle, the long cycle utilizes a predictor-corrector scheme with step size controlled by the maximum local error in orbital parameters $\epsilon_{\rm{L}}$. Depending on the planet's rheology and loading frequency, the long step size may gradually become several orders of magnitude larger than the step size of the short cycle. This combined integration scheme allows us to take relatively large steps in the evolution of orbital parameters, while still keeping precise value of the current spin state.

During the despinning mode of the short cycle, as well as in each step of the long cycle, we evolve also the planet's thermal state and interior structure. For the sake of clarity, let $\Delta t_{\rm{orb}}$ symbolize the current step size in either of the orbital evolution cycles from which we call the convection module. Equations (\ref{eq:tempeqM}) and (\ref{eq:tempeqC}), controlling the interior temperature, and equation (\ref{eq:del_lid}), which describes the evolution of the stagnant lid, are solved explicitly with a step size equal to $\min(\Delta t_{\rm{orb}},\unit[10^4]{yr})$. As the interior temperature evolves, we actualize also the mantle viscosity and rigidity and inspect whether the planet contains a magma ocean.\\

Since our model consists of two parts---a tidal module and a convection module---we seek the mantle viscosity and rigidity in two different forms. In order to calculate the mean Rayleigh number, which is used for the calculation of the upper thermal boundary thickness, we require mantle viscosity averaged over the entire mantle. The tidal model, on the other hand, enables us to divide the planetary mantle into several layers and endow each of them with its own, locally-averaged rheological parameters. Specifically, in the case of a strongly tidally heated planet, the mantle can be divided into a solid region, with zero or relatively low melt fraction ($\phi<\phi_{\rm{D}}$), and a magma ocean, in which the melt fraction exceeds the disaggregation point ($\phi>\phi_{\rm{D}}$).

The viscosity and the rigidity are calculated in different depths of the mantle, following relations (\ref{eq:muphi}) and (\ref{eq:etaphi}) and assuming adiabatic temperature profile. The average mantle viscosity $\bar{\eta}_{\rm{m}}$, which enters the convection module, is then obtained as the geometric mean of the radially dependent values. At the same time, we inspect the radially dependent melt fraction (\ref{eq:meltfr}). If it never exceeds the disaggregation point, the mantle is solid and the tidal model consists of a single mantle layer, whose viscosity is equal to $\bar{\eta}_{\rm{m}}$. Conversely, if the melt fraction at any radius $r$ exceeds the disaggregation point, the model should contain a magma ocean. In this case, we seek the lower boundary of the ocean and divide the mantle in the tidal model into two layers. The average viscosity and rigidity of these layers are then calculated from the radially dependent quantities individually.

Although the parametrized mantle convection model includes the evolution of the stagnant lid thickness, we do not consider melting of the planet's surface. Once the lid thickness decreases to a prescribed minimum value $D_{\rm{l, min}}$, it is set constant. The lithosphere is then allowed to evolve only after the planet begins to cool down. Similarly, in case the temperature at the bottom of the convecting mantle $T_{\rm{b}}$ exceeds the temperature at the core-mantle boundary $T_{\rm{c}}$, we set the thickness of the lower boundary layer to a constant value $\delta_{\rm{c, par}}$. The heat flux from the mantle to the core is then calculated using equation (\ref{eq:qc}) with the parameter $\delta_{\rm{c, par}}$ instead of $\delta_{\rm{c}}$.

A list of the numerical parameters used in this study is given in Table \ref{tab:Nparams}.

\begin{table}[h]
\caption{Parameters of the numerical scheme}\label{tab:Nparams}
\begin{center}
\begin{tabular}{l l l}
\hline
Parameter              & Definition & Value or interval  \\
\hline
$\epsilon_{\rm{S}}$ & Local error in the short cycle & $[10^{-12},10^{-10}]$ or lower \\
$\epsilon_{\rm{L}}$ & Local error in the long cycle & $[10^{-10},10^{-8}]$ \\
$\delta_{\rm{S}}$ & Upper limit on $\ddot{\theta}/\dot{\theta}$ in the short cycle& $10^{-16}$ \\
\hline
\end{tabular}
\end{center}
\end{table}

\section{Tidal heating and the highest stable spin-state} \label{sec:results1}

In this section, the first of the sections dedicated to the results of our study, we are going to investigate the effect of rheological and orbital parameters on the tidal dissipation and tidal locking of a generic terrestrial exoplanet hosted by a red dwarf (an M-type star). Planetary systems around M-type stars are a particularly interesting target. Their conventional habitable zone lies very close to the primary and, therefore, overlaps with the region of strong tidal loading \citep[e.g.,][]{behounkova11}. Knowledge of the thermal and orbital state of such bodies may serve as an additional information for the evaluation of planetary habitability \citep[e.g.,][]{wandel18,godolt19}.

The secular thermal and orbital evolution of close-in exoplanets is determined by the rate of the energy dissipation and by the planet's spin rate. Furthermore, the rotation state in which the planet resides also affects its surface conditions and, potentially, the evolution of the climate and habitability prospects. Tidal locking into synchronous rotation with its extreme insolation pattern results in essentially different climate forcing than faster, nonsynchronous rotation \citep[e.g.,][]{dobro07}. To illustrate the role of various parameters on the tidal dissipation and on the spin rate, we first perform several numerical studies without taking into account the planet's internal dynamics, variations of orbital elements, and changes of the interior structure. Throughout this section, the only variables evolving in time are the spin rate and the surface tidal heat flux.\\

Our model planet consists of three layers: a liquid core with low, finite viscosity, a viscoelastic mantle governed by the Andrade rheology, and an elastic lithosphere of constant thickness. In the parametric studies, we do not explicitly include the magma ocean as a separate layer. The rheological properties of the planet, as well as other model parameters, are listed in Table \ref{tab:Aparams}. Among other, previously introduced parameters, Table \ref{tab:Aparams} includes the core mass fraction $\rm{CMF}$, which is defined as the ratio of the core mass to the total mass of the planet. Top radii of the core and of the mantle with given densities are chosen to match the total radius and the given core mass fraction.

Throughout the parametric studies, the planet is first allowed to despin from arbitrarily chosen initial spin-orbit ratio $\dot{\theta}/n=5.6$ to the first (i.e., highest) stable spin state, in which the derivative $\ddot{\theta}$ decreases below $\delta_{\rm{S}}$ (see Table \ref{tab:Nparams}). Tidal heating in the equilibrium spin state is then calculated by formula (\ref{eq:segatz_final}). In order to normalize the tidal heat rate to the planet's surface, we introduce the average surface tidal heat flux,

\footnotesize
\begin{equation}
    \Phi_{\rm{tide}} = \frac{\bar{P}^{\rm{tide}}}{4\pi R^2}\; ,
\end{equation}\normalsize\\
which facilitates the comparison with total heat flux at the surface of the Earth ($\unit[0.09]{mW\;m^{-2}}$; \citealp{davies10}) or Io ($>\unit[2.5]{W\;m^{-2}}$; \citealp{veeder94}).

\begin{table}[h]
\caption{Parameters of the generic terrestrial exoplanet}\label{tab:Aparams}
\begin{center}
\begin{tabular}{l l l l}
\hline
Parameter & Definition & Value & Unit  \\
\hline
$m_*$ & Mass of the host star & $0.1$ & $m_{\sun}$  \\
$a$ & Semi-major axis & $0.04$ & $\unit{AU}$  \\
$e$ & Eccentricity & $0.0$ to $0.5$ & ---  \\
\hline
$\rho_{\rm{c}}$ & Core density & $9000$ & $\unit{kg\; m^{-3}}$  \\
$\eta_{\rm{c}}$ & Core viscosity & $10^{-3}$ & $\unit{Pa\, s}$  \\
$\mu_{\rm{c}}$ & Core rigidity & $10^{-10}$ & $\unit{Pa}$  \\
\hline
$\rho_{\rm{m}}$ & Mantle density & $5000$ & $\unit{kg\; m^{-3}}$  \\
$\eta_{\rm{m}}$ & Mantle viscosity & $10^{10}$ to $10^{22}$ & $\unit{{Pa\, s}}$  \\
$\mu_{\rm{m}}$ & Mantle rigidity & $10^{6}$ to $10^{15}$ & $\unit{Pa}$ \\
\hline
$\rho_{\rm{lid}}$ & Lithosphere density & $3000$ & $\unit{kg\; m^{-3}}$  \\
$\mu_{\rm{lid}}$ & Lithosphere rigidity & $7\times 10^{10}$ & $\unit{Pa}$  \\
\hline
$d_{\rm{lid}}$ & Lithospfere thickness & $50$ & $\unit{km}$  \\
$\rm{CMF}$ & Core mass fraction & $0.1$ to $0.7$ & ---  \\
$R$ & Outer radius of the planet & $0.2$ to $1.5$ & $R_{\earth}$ \\
$\alpha$ & Parameter of the Andrade model & $0.3$ & ---  \\
$\zeta$ & Parameter of the Andrade model & $1$ & ---  \\
\hline
\end{tabular}
\end{center}
\end{table}

\subsection{Effect of rheological parameters} \label{sub:etamu}

\begin{figure}[h]
  \begin{center}
    \includegraphics[width=1.\textwidth]{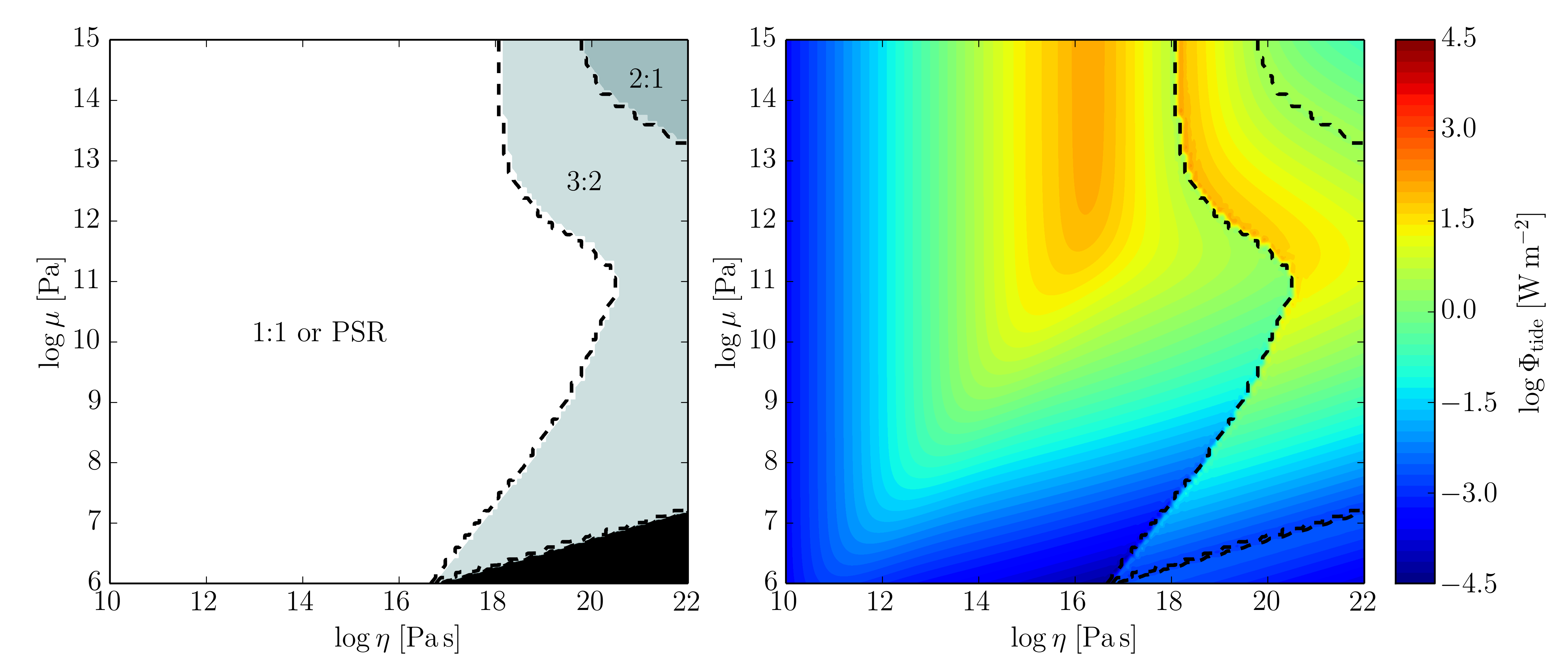}
  \end{center}
\caption{The highest stable spin-state (\textit{left}) and the surface tidal heat flux (\textit{right}) of the model planet with orbital eccentricity $e=0.05$. The caption "PSR" corresponds to pseudosynchronous rotation with spin-orbit ratio $\dot{\theta}/n\approx1.015$. Triangular region in the lower right corner of both panels indicates combinations of parameters for which the tidal despinning takes more than $\unit[1]{Gyr}$.} \label{fig:etamu_e005}
\end{figure}

\begin{figure}
  \begin{center}
    \includegraphics[width=1.\textwidth]{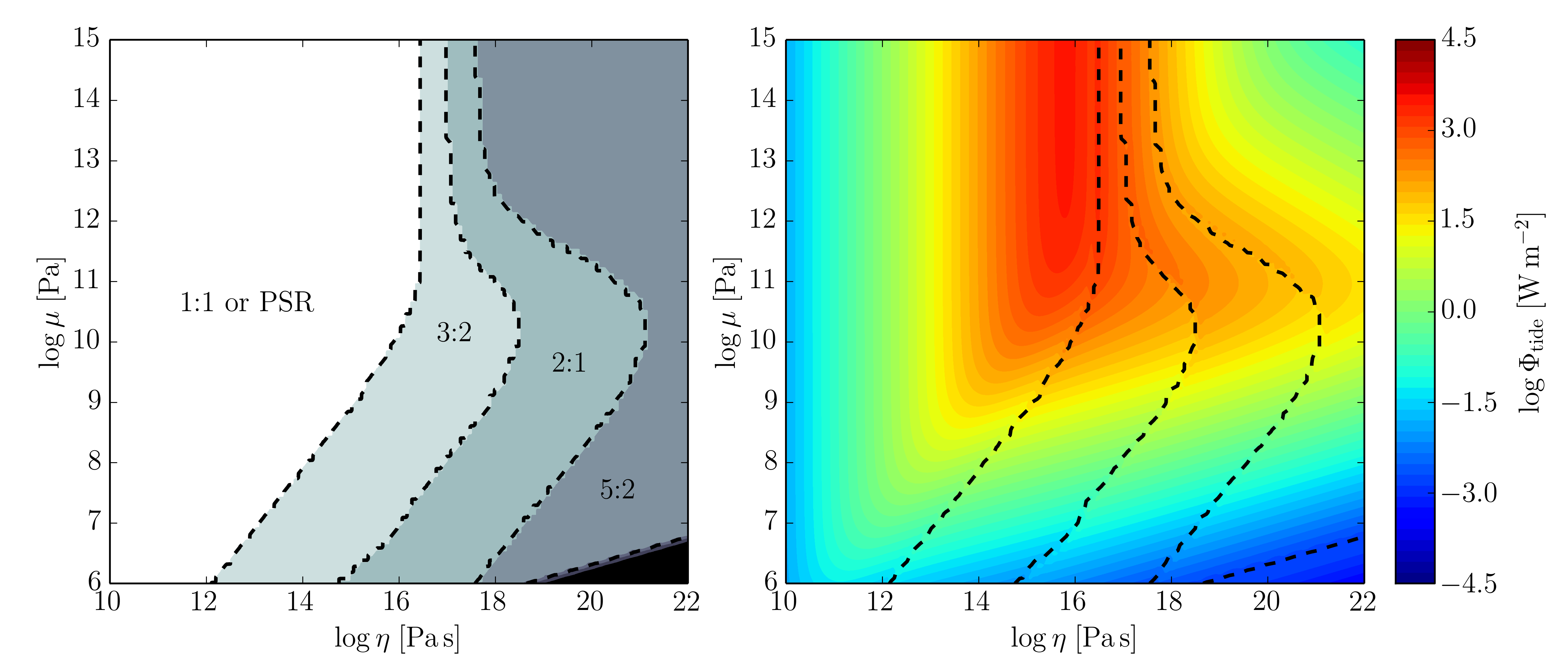}
  \end{center}
\caption{Same as Figure \ref{fig:etamu_e005}, but for orbital eccentricity $e=0.2$. Pseudosynchronous rotation (PSR) corresponds to $\dot{\theta}/n\approx1.24$.} \label{fig:etamu_e02}
\end{figure}

\begin{figure}
  \begin{center}
    \includegraphics[width=1.\textwidth]{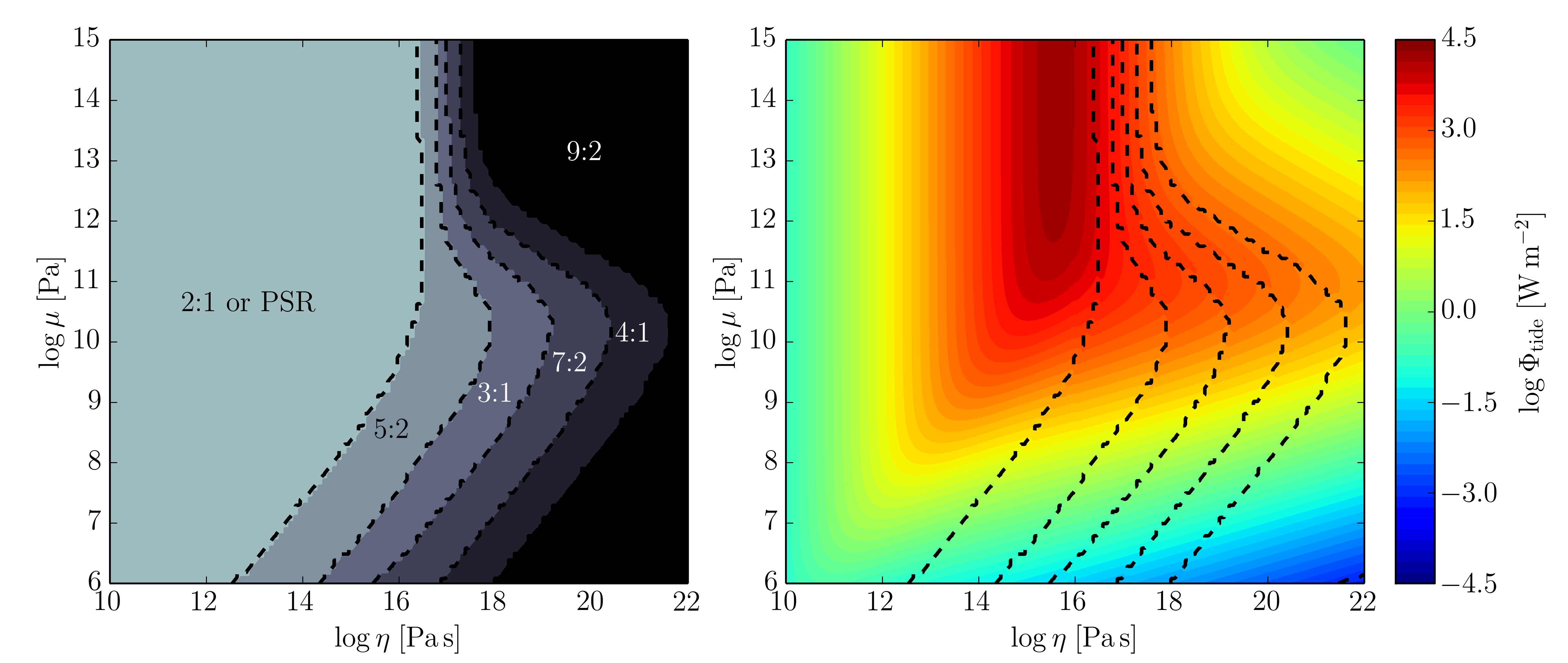}
  \end{center}
\caption{Same as Figure \ref{fig:etamu_e005}, but for orbital eccentricity $e=0.4$. Pseudosynchronous rotation (PSR) corresponds to $\dot{\theta}/n\approx1.96$.} \label{fig:etamu_e04}
\end{figure}

In this subsection, we consider an Earth-sized model planet ($R=1R_{\earth}$) with Earth-like core mass fraction ($\rm{CMF}=0.3$) and investigate the effect of varying mantle rigidity and viscosity for three different orbital eccentricities: $0.05$, $0.2$ and $0.4$. Figures \ref{fig:etamu_e005}-\ref{fig:etamu_e04} depict the regions of the parametric space with different highest stable spin states and the corresponding surface tidal heat fluxes. Although in most of the cases the model planet despins into a spin-orbit resonance, we note that at very low mantle viscosities the stable spin state is pseudosynchronous, with spin-orbit ratio approximately given by \citep[e.g.,][]{dobro07}

\footnotesize
\begin{equation}
\frac{\dot{\theta}}{n} \approx 1 + 6e^2\; .
\end{equation}\normalsize\\
Nevertheless, in the illustrations we include the region of pseudosynchronous rotation ("PSR") into the region of the closest spin-orbit resonance.

A common feature of all the model cases is the complex shape of boundaries between different stable spin states. While for high mantle rigidities ($>\unit[10^{12}]{Pa}$) the first stable spin state depends almost exclusively on the viscosity, for low mantle rigidities ($<\unit[10^{10}]{Pa}$) it depends on both of the rheological parameters. If evolving from this region, a planet with initially low mantle rigidity would be more susceptible to tidal locking into a high spin-orbit resonance, which could be eventually destabilized by an increase in the rigidity at constant viscosity. The transition between the two tendencies in tidal locking is characterized by a heap in the boundary between different high spin-orbit resonances. Described behaviour can be understood by investigating the parameter dependence of tidal torque or angular acceleration (\ref{eq:rot}), as is done in Appendix \ref{app2} for the Maxwell model. Specifically, the low-rigidity case corresponds to a self-gravity-dominated deformation regime, while the high-viscosity case is governed by the planet's rheology.

Another pattern observable in Figures \ref{fig:etamu_e005}-\ref{fig:etamu_e04} is the eccentricity dependence of width of the regions with different highest stable spin state. At high orbital eccentricity ($e=0.4$) and for an arbitrarily chosen rigidity of $\unit[10^{12}]{Pa}$, the change in the mantle viscosity from $10^{18}$ to $\unit[10^{16}]{Pa\,s}$ would result in a steep cascade of spin state transitions. For less eccentric orbits, on the other hand, the evolution of spin rate would be much more gradual.\\

The effect of the mantle rigidity and viscosity on the surface tidal heat flux of synchronously rotating moons and exoplanets has already been discussed in an extensive literature \citep[e.g.,][]{fischer90,moore03,henning09,renaud18}. Here, we only mention that the role of the mantle rigidity and viscosity is considerably stronger than the effect of different spin states and orbital eccentricities. However, in the model cases with low orbital eccentricity, the surface tidal heat flux apparently depends on the spin-orbit ratio (Figure \ref{fig:etamu_e005}). A transition between the $3:2$ spin-orbit resonance and the synchronous rotation may result in order of magnitude drop in the surface tidal heat flux, which would be probably succeeded by a significant change in the surface and interior conditions.

\subsection{Effect of orbital eccentricity}

In the second parametric study, we set the mantle rigidity to a constant value of $\mu_{\rm{m}}=\unit[200]{GPa}$ and vary the mantle viscosity and the orbital eccentricity. The remaining parameters are the same as in the previous section. To assure the precision of the computation, we chose the cut-off degree of the Kaula's eccentricity functions $\mathcal{G}_{lpq}(e)$ with respect to the actual eccentricity. Specifically, we required that the truncation error of the disturbing potential and evolution equations was lower than $10^{-4}$ and we continued to increase the upper limit for index $q$ of the Darwin-Kaula expansion from $q_{\rm{max}}=1$ for the lowest eccentricities up to $q_{\rm{max}}=7$ for $e=0.5$.

\begin{figure} 
  \begin{center}
    \includegraphics[width=1.\textwidth]{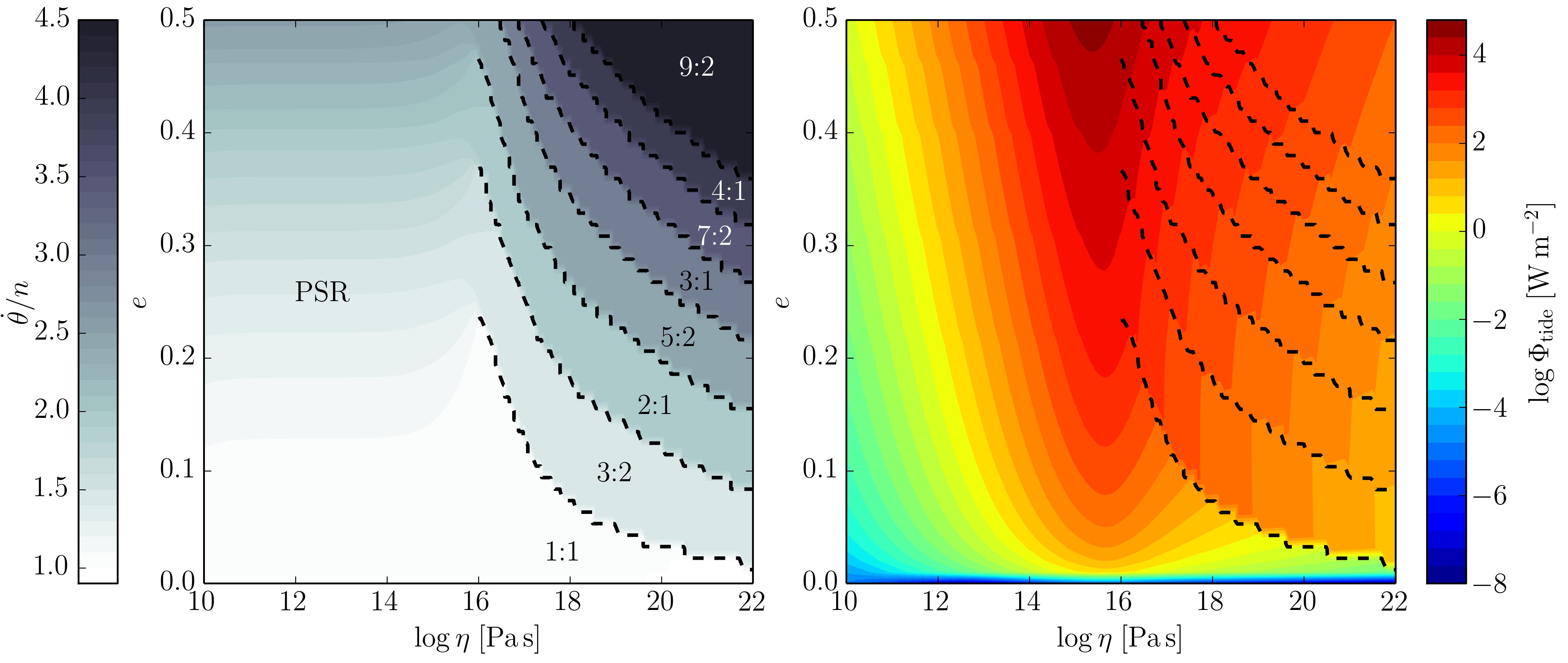}
  \end{center}
\caption{The highest stable spin-state (\textit{left}) and the surface tidal heat flux (\textit{right}) for a model planet with mantle rigidity $\mu_{\rm{m}}=\unit[200]{GPa}$. Caption "PSR" corresponds to pseudosynchronous rotation.} \label{fig:eta_ecc}
\end{figure}

Figure \ref{fig:eta_ecc} unveils two distinct regions, which are characterized by a different type of stable spin states and different parameter dependence of tidal dissipation. The boundary between these two regions is due to the change in behaviour of the viscoelastic material at different tidal frequencies, which can be related to its characteristic (Maxwell) time $\tau_{\rm{M}}={\eta_{\rm{m}}}/{\mu_{\rm{m}}}$. In general, planets are loaded on a variety of frequencies; however, for the simplest case of a synchronously rotating body on a slightly eccentric orbit, the tidal frequency is approximately equal to the orbital frequency. Since the orbital parameters of our model system yield $T_{\rm{orb}}=\unit[9.24]{days}$, it follows that planets with mantle viscosity higher than $\unit[10^{17}]{Pa\,s}$ are loaded on periods shorter than their Maxwell time and behave more viscoelastically, while planets with considerably lower mantle viscosities can be considered as purely viscous.

The first of the two regions, at low mantle viscosities, is, therefore, characterized by pseudosynchronous rotation and by tidal dissipation which smoothly increases with increasing orbital eccentricity or proximity to the boundary between the zones (below $\eta_{\rm{m}}\approx\unit[10^{16}]{Pa\,s}$). In this region, the surface tidal heat flux is determined primarily by the mantle viscosity, with a comparably weaker contribution of the orbital eccentricity. On the contrary, the rotational evolution of a solid body lying in the second (high-viscosity) region is marked by transitions between stable spin-orbit resonances. Their stability is given predominantly by the orbital eccentricity, with more eccentric orbit resulting in higher first stable resonance, but it can be also affected by changing the viscosity, as was the case in the previous section.

Looking at the tidal dissipation, the average surface tidal heat flux of a planet locked in a given spin-orbit resonance is only weakly dependent on the orbital eccentricity and changes mainly due to the variations in the viscosity \citep[see also][]{behounkova11}, with the exception of the transition to synchronous rotation. This behaviour is very different from the viscous region and can be derived from the expression for the average tidal heat rate\footnote{The average tidal heat rate (\ref{eq:segatz_final}), written in the form of the Darwin-Kaula expansion, contains products of the eccentricity polynomials $\mathcal{G}_{lpq}(e) \in o(e^{|q|})$. Since we restricted our study to the case of zero obliquity, the only nonzero terms in the expansion are $\{lmpq\}=\{220q\}$ and $\{lmpq\}=\{201q\}$, for which the higher the index $q$, the weaker the contribution of the term to the total sum. Specifically, the term with $q=0$ is independent of the eccentricity. Each term of the expression (\ref{eq:segatz_final}) is also multiplied by the tidal frequency $\omega_{lmpq}$, which can be in our case either $\omega_{220q}=(2+q)n-2\dot{\theta}$ or $\omega_{201q}=qn$. For a synchronously rotating body ($\dot{\theta}=n$), both of the considered frequencies are zero for $q=0$, and the leading term is, therefore, $q=1$. For a higher spin-orbit resonance, the frequency $\omega_{2200}$ is nonzero and the leading term does not depend on the eccentricity.}.

\subsection{Effect of interior structure}

In order to investigate the role of the planet's radius and core mass fraction on the stability of spin-orbit resonances and on the tidal heating, we again set the mantle rigidity to a fixed value $\mu_{\rm{m}}=\unit[200]{GPa}$, as in the previous section. Additionally, we consider Earth-like value of the mantle viscosity $\eta_{\rm{m}}=\unit[10^{21}]{Pa\, s}$ and two possible values of orbital eccentricity, $e=0.05$ or $e=0.2$. The densities of all interior layers are considered constant, as listed in Table \ref{tab:Aparams}.

\begin{figure}
  \begin{center}
    \includegraphics[width=1.\textwidth]{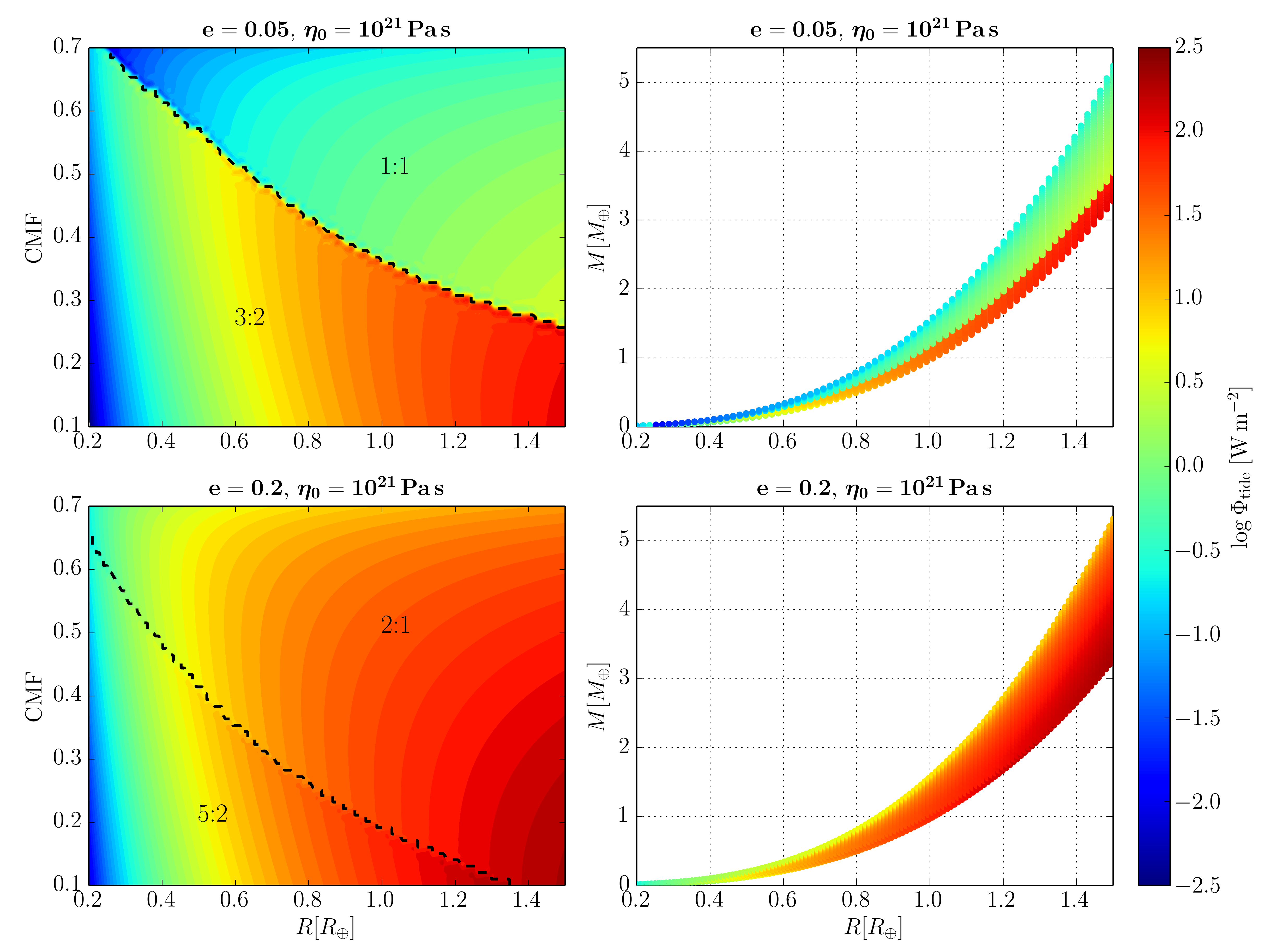}
  \end{center}
\caption{Surface tidal heat flux as a function of the planet's radius and core mass fraction (\textit{left}) and the same figure plotted into a mass-radius graph (\textit{right}). Dashed line demarcates the boundaries between regions with different highest stable spin state. The mantle viscosity was set to $\eta_{{\rm{m}}}=\unit[10^{21}]{Pa\,s}$ and the orbital eccentricity to $e=0.05$ (\textit{upper row}) or $e=0.2$ (\textit{lower row}).} \label{fig:rcmf_eta21}
\end{figure}

Figure \ref{fig:rcmf_eta21} depicts the average surface tidal heat flux with inscribed boundaries between different highest stable spin states, as well as a simplified mass-radius diagram of the model ensemble. Focusing first on the spin rate evolution, we see that the main feature of the results is higher susceptibility of planets with small radii or low core mass fractions to tidal locking into higher spin-orbit resonances. This observation also means that the lower the planet's mass for a given radius, the higher the probability that the planet rotates nonsynchronously. For higher orbital eccentricities, the model planet gets always locked into higher than synchronous resonances; however, the effect of the small radius or low core mass fraction on the spin-orbit ratio remains qualitatively the same.

Again, the planet's spin state cooperates with other model parameters on determining the rate of energy dissipation. Similarly to the previous results, we may see that the effect of tidal locking on the tidal heating is most prominent in the case of planets with low orbital eccentricity (upper row in Figure \ref{fig:rcmf_eta21}), where trespassing the boundary between the synchronous and nonsynchronous rotation results in two orders of magnitude change in the surface tidal heat flux. A common feature of all model cases is increasing surface tidal heat flux with increasing planetary radius or decreasing core mass fraction, which is a simple consequence of differing mantle volume. Translated into the mass-radius diagram, the higher the mass of a nonsynchronously rotating planet of a given radius, the lower the tidal heat flux, given that the rheological parameters remain constant.

\section{Application to low-mass exoplanets} \label{sec:results2}

Extrasolar planets with radii below $1.5\; R_{\earth}$ \citep{weiss14} or masses below $2-4\; M_{\earth}$ \citep{chen17} are expected to have rocky composition, similarly to the terrestrial worlds of the Solar System. At the same time, many of these \textit{terrestrial exoplanets} orbit very close to their host star and their thermal and orbital evolution has been presumably marked by a period of substantial tidal dissipation. Since the theoretically predicted final state of a tidally evolved exoplanet is a circular orbit and synchronous rotation, attention has been recently drawn to the number of bodies whose orbital eccentricity seems to be, despite their proximity to the host star, still nonzero.

Among other explanations, such as observational bias, low age of the system, gravitational scattering or eccentricity excitation by mean motion resonances, it has been proposed \citep[e.g.,][]{henning09,makarov15} that the nonzero eccentricities can be maintained by the thermal state of the planet. This mechanism has been illustrated in several studies \citep[e.g.,][]{henning09,shoji14,driscoll15} and can be also deduced from Figures \ref{fig:etamu_e005}-\ref{fig:etamu_e04}. If a rocky planet with initial mantle viscosity $\unit[10^{21}]{Pa\, s}$ and rigidity $\unit[200]{GPa}$ begins to melt, the decrease in viscosity leads it close to the region of maximum, runaway heating, which is, in our case, around $\unit[10^{16}]{Pa\, s}$. As a consequence of the increased heat generation, the lattice of mantle minerals disrupts and the mantle begins to melt. At this stage, the mantle viscosity and rigidity decrease abruptly and terminate the period of extreme tidal heating. Furthermore, the change in the rheological parameters might also result in change of the planet's spin state \citep{makarov18}.\\

In this section, we are going to perform the parametric study of rheological properties for the models of three low-mass exoplanets: GJ 625 b \citep{suarez17}, GJ 411 b \citep{diaz19}, and Proxima Centauri b \citep{anglada-escude16}. As in Section \ref{sec:results1}, we explore here only the effect of instantaneous rheological and orbital parameters on the tidal locking and dissipation; i.e., we consider neither the thermal nor the orbital evolution. The planets have been chosen on the grounds of their masses, proximities to the host star, nonzero eccentricities and presumed absence of strong perturbations by other bodies in the system. Since all of these exoplanets were found by radial velocity measurements and only their minimum masses are known, we calculate the (minimum) radii from the mass-radius relation of \citet{zeng16},

\footnotesize
\begin{equation}
\left(\frac{R}{R_{\earth}}\right) = (1.07 - 0.21\cdot{\rm{CMF}}) \left(\frac{M}{M_{\earth}}\right)^{{1}/{3.7}}
\end{equation}\normalsize\\
and assume that they have an Earth-like core mass fraction $\rm{CMF}=0.3$. The average core density is set to $\rho_{\rm{c}}=\unit[10000]{kg\,m^{-3}}$ in the case of Proxima Centauri b and to $\rho_{\rm{c}}=\unit[12000]{kg\,m^{-3}}$ for the other two exoplanets and the respective average mantle densities are calculated to match the prescribed masses and radii. Numerical values of all parameters used for the following study are listed in Table \ref{tab:Bparams}.

\begin{table}[h]
\caption{Model parameters of the studied exoplanets}\label{tab:Bparams}
\begin{center}
\begin{tabular}{l l l l}
\hline
Parameter & Proxima Centauri b & GJ 625 b & GJ 411 b  \\
\hline
$m_*\; [m_{\sun}]$ & $0.12$ & $0.30$ & $0.39$ \\
$a\; [\unit{AU}]$ & $0.0485$ & $0.0784$ & $0.0785$ \\
\hline
$m_{\rm{p}}\; [m_{\earth}]$ & $1.27$ & $2.82$ & $2.99$\\
$R\; [R_{\earth}]$ & $1.074$ & $1.333$ & $1.354$ \\
$\rho_{\rm{c}}\; [\unit{kg\;m^{-3}}]$ & $10000$ & $12000$ & $12000$ \\
$\rho_{\rm{m}}\; [\unit{kg\;m^{-3}}]$ & $4797$ & $5502$ & $5589$ \\
\hline
\end{tabular}
\end{center}
\end{table}

\subsection{GJ 625 b}

Exoplanet GJ 625 b is about three times more massive than the Earth and lies on the inner edge of the habitable zone of its host star \citep{suarez17}. Depending on its cloud coverage and albedo, it might or might not be able to support the existence of liquid surface water. Although no other planets have been discovered in the system, the orbital eccentricity of GJ 625 b was estimated as $e=0.13\substack{+0.12 \\ -0.09}$ \citep{suarez17}, which may point out at the effect of other mechanisms than eccentricity excitation by mean motion resonance.

\begin{figure}[ht]
  \begin{center}
    \includegraphics[width=1.\textwidth]{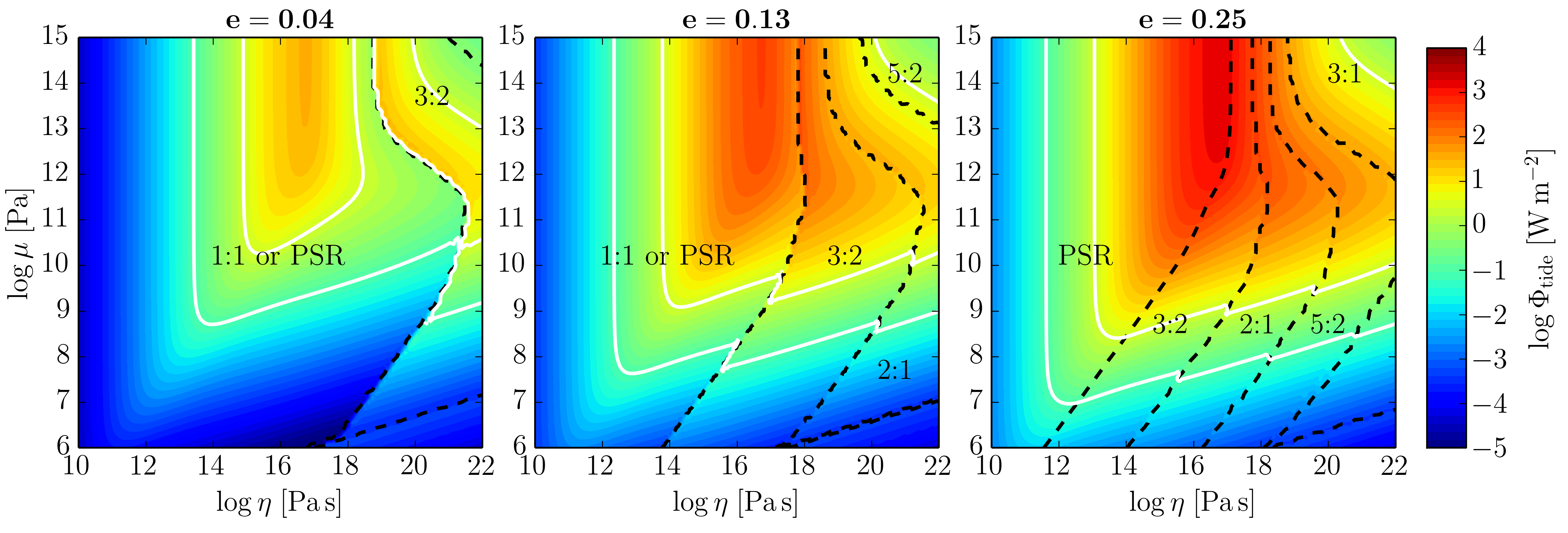}
  \end{center}
\caption{Surface tidal heat flux of a model of GJ 625 b for three plausible orbital eccentricities. Dashed lines delimit boundaries between the regions with different highest stable spin state. Solid white lines indicate the total surface heat flux on the Earth ($\sim\unit[0.09]{W\;m^{-2}}$) and on Io ($\sim\unit[2.5]{W\;m^{-2}}$). Triangular region in the lower right corners of the panels indicates combinations of parameters for which the despinning takes more than $\unit[10]{Gyr}$.} \label{fig:gj625_hf}
\end{figure}

Figure \ref{fig:gj625_hf} shows the effect of the mantle viscosity and rigidity on the planet's spin state and surface tidal heat flux. The parametric study is performed for the mean and limit values of orbital eccentricity, that is, $e=0.04$, $0.13$, and $0.25$. Considering, for the sake of illustration, a reference Earth-like viscosity $\eta_{\rm{m}}=\unit[10^{21}]{Pa\, s}$ and rigidity $\mu_{\rm{m}}=\unit[200]{GPa}$, we see that the planet always despins into higher than synchronous spin-orbit resonance and retains significant surface tidal heat flux. Since the tidal heat flux for all studied values of eccentricity exceeds the surface heat flux at Io, the planet in this model settings is not expected to be habitable, independently of the incident flux from the host star. A more favorable situation would arise if the average mantle viscosity and rigidity were reduced either due to a different mineralogical composition or due to the presence of subsurface melt or water. In the case of, e.g., $\eta_{\rm{m}}=\unit[10^{19}]{Pa\, s}$ and $\mu_{\rm{m}}=\unit[100]{MPa}$, the surface tidal heat flux becomes comparable to the total heat flux at the Earth and the tidal effects do not present substantial threat to potential habitability. Only in the case of the lowest considered eccentricity ($e=0.04$) does this combination of rheological parameters result in tidal locking into the synchronous rotation, whose effect on the surface conditions is ambiguous \citep[e.g.][]{kite11,checlair19}.

\subsection{GJ 411 b}

Red dwarf GJ 411 belongs to the closest stars from the Sun and it is also one of the brightest M-dwarfs on the Earth sky \citep{lepine11}. As for December 2019, it is known to harbour one confirmed exoplanet, whose mass might be compatible with rocky composition \citep{diaz19}. The planet, GJ 411 b, orbits the star with a $12.95$-day period and its equilibrium surface temperature lies between $\unit[256]{K}$ and $\unit[350]{K}$, depending on the albedo. It is, therefore, not expected to be habitable. Furthermore, its presumably nonzero eccentricity $e\in[0.00,0.44]$, with most likely value of $e=0.22\pm0.13$, \citep{diaz19} makes it susceptible to immense tidal loading.

\begin{figure}[ht]
  \begin{center}
    \includegraphics[width=1.\textwidth]{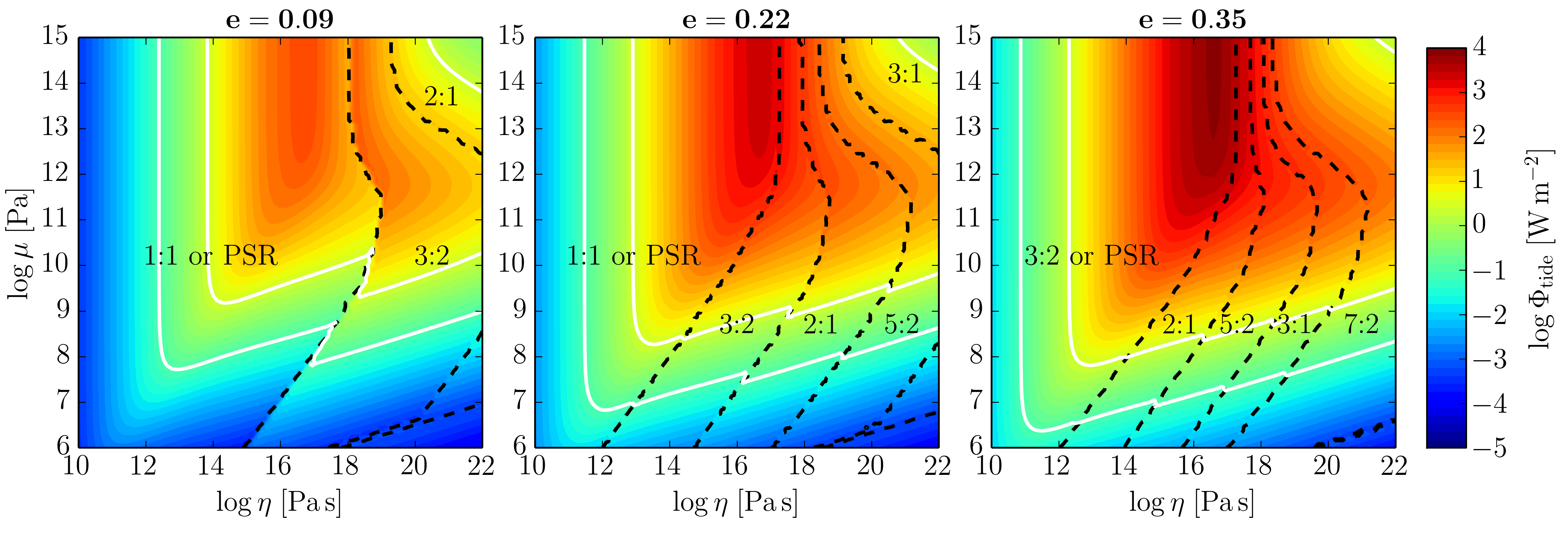}
  \end{center}
\caption{Same as Figure \ref{fig:gj625_hf}, but for GJ 411 b.} \label{fig:gj411_hf}
\end{figure}

Figure \ref{fig:gj411_hf} depicts the role of rheological parameters in the thermal-rotational evolution of GJ 411 b. Since the considered eccentricities are relatively high, the planet with a reference Earth-like composition would be most probably locked in a spin-orbit resonance between $3:2$ and $3:1$. Partial melting, which generally sets the planet to the central (red) part of the graph, would lead, in the case of small eccentricity, to pseudosynchronous rotation. For higher eccentricities, however, even a molten planet with low mantle rigidity and decreased viscosity keeps a resonant spin state. The pattern of tidal heating is very similar to the previous case and the surface tidal heat flux for realistic combinations of rheological parameters again exceeds the surface heat flux at Io. Combined with the high insolation, the tidal dissipation may contribute to transforming GJ 411 b into a lava planet.

\subsection{Proxima Centauri b}

The discovery of an Earth-mass planet on the orbit around Proxima Centauri \citep{anglada-escude16} has drawn a lot of attention mainly due to its astrobiological significance. Not only does the planet dwell in the traditional habitable zone, but its proximity makes it also a suitable target for direct imaging in the near future \citep{turbet16}. The orbit of Proxima Centauri b indicates a small remnant eccentricity of $e=0.08\substack{+0.07 \\ -0.06}$ \citep{jenkins19} and, depending on the efficiency of tidal dissipation, the planet is expected to be locked either in the state of synchronous rotation or in the $3:2$ spin-orbit resonance \citep{ribas16}. In a recent analysis of new radial velocity data for the system, \citet{damasso20} found that Proxima Centauri may also host another low-mass exoplanet on a wide orbit. However, since the orbital period of the new exoplanet candidate is two orders of magnitude higher than the orbital period of Proxima Centauri b, it is not expected to affect the dynamics of the latter significantly.\\

\begin{figure}[ht]
  \begin{center}
    \includegraphics[width=1.\textwidth]{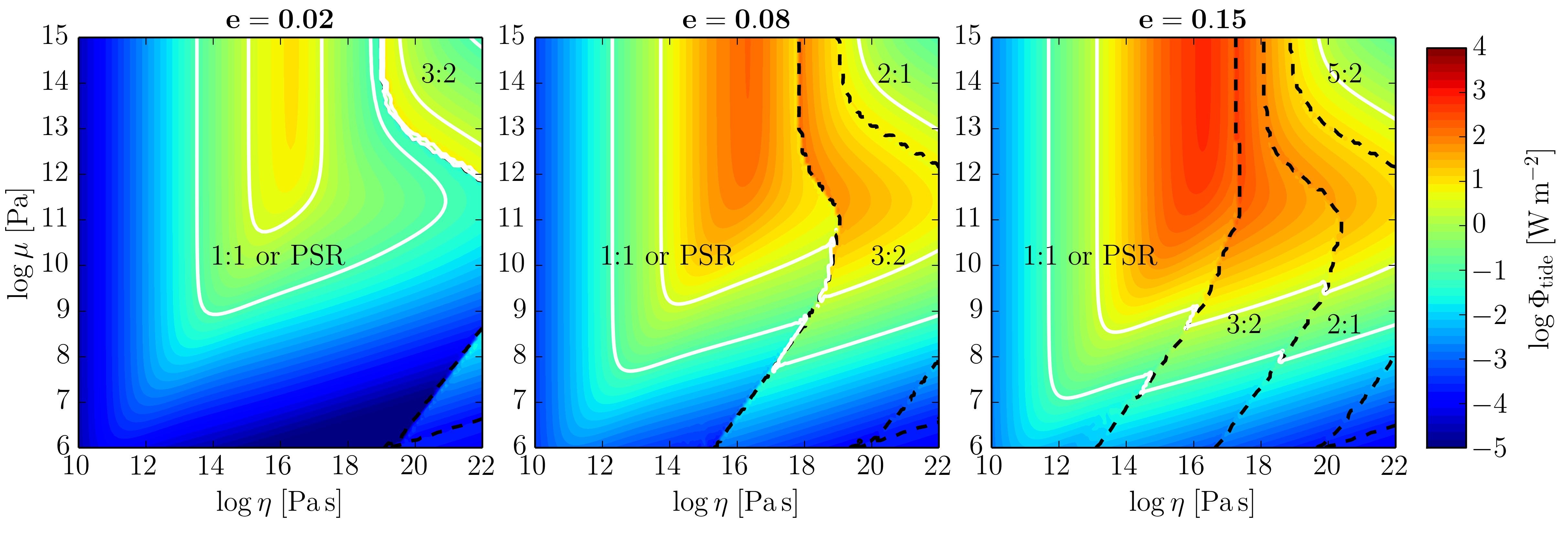}
  \end{center}
\caption{Same as Figure \ref{fig:gj625_hf}, but for Proxima Centauri b.} \label{fig:proxima_hf}
\end{figure}

The results of our last parametric study with constant parameters are illustrated in Figure \ref{fig:proxima_hf}. In accordance with the study of \citet{ribas16}, the planet on the least eccentric orbit tends to the synchronous rotation for most of the considered pairs of rheological parameters. Only the combination of Earth-like viscosity ($>\unit[10^{20}]{Pa\, s}$) and very low ($<\unit[10^{8}]{Pa}$) or high ($>\unit[10^{12}]{Pa}$) rigidity may allow the preservation of the $3:2$ spin-orbit resonance even on almost circular orbit. In the case of low rigidity, however, the tidal dissipation is almost negligible, resulting in a rather long period of despinning (on the scale of billions of years) and, in the most extreme situation, even preventing the planet from reaching stable spin state during the first $\unit[10]{Gyr}$. For the other two considered eccentricities, the planet with Earth-like parameters despins first into higher spin-orbit resonance ($3:2$ or $2:1$) and is able to maintain surface heat flux much higher than Io, which would later entail its melting and further despinning. In the view of the parametric study with fixed interior, we conclude that the habitability prospects of Proxima Centauri b are strongly dependent on the orbital eccentricity and for even mildly eccentric orbits they may be limited.

\section{Coupled thermal-orbital evolution} \label{sec:results3}

Previous sections have shown that tidal heating is a complex function of various orbital and rheological parameters. The same complexity is also reflected by other phenomena affected by energy dissipation. Namely, it marks the rate of orbital evolution and the vigour of mantle convection. Stepping away from the parametric studies with constant orbital elements and interior structure, we are ready to discuss the results of the fully coupled model with emerging magma ocean.

The following two subsections are dedicated to the long-term evolution of the exoplanets described in Table \ref{tab:Bparams}. However, since we are now considering also the thermal evolution, it is necessary to include several new quantities to the analysis, such as the initial temperatures and the parameters of the convection model (Table \ref{tab:Cparams}). In all model cases, we assume that the planet's mantle is initially solid and the temperature at the top of the convective layer is $T_{\rm{m, 0}}=\unit[1500]{K}$. The tidal evolution of real exoplanets would presumably initiate in the magma ocean state, when at least part of the planet is still molten due to the heat released during its formation. The interior temperatures would be, therefore, much higher. To account for these initial conditions, we also conducted numerical experiments for initial temperatures $T_{\rm{m, 0}}=\unit[2000]{K}$ and $T_{\rm{m, 0}}=\unit[2500]{K}$; however, all cases led to a rapid equilibration of $T_{\rm{m}}$ around the same value. The effect of the initial temperature at the top of the convecting mantle is, in the long term, minimal. We also tested different values of the disaggregation widths $\Delta_{\eta}$ and $\Delta_{\mu}$, which affect the transition of a partially molten layer from a solid-like behaviour to a liquid-like behaviour around the disaggregation point $\Phi_{\rm{D}}$. Higher $\Delta$ corresponds to a more gradual transition between the two regimes. On the base of these numerical experiments, we chose the value $\Delta_{\eta}=\Delta_{\mu}=0.01$, since it best describes the abrupt change considered in other studies \citep[e.g.,][]{fischer90} and since further decrease in $\Delta$ does not substantially affect the results.

\begin{table}[h]
\caption{Parameters of the mantle convection model}\label{tab:Cparams}
\begin{center}
\begin{tabular}{l l l l}
\hline
Parameter & Definition & Value & Unit \\
\hline
$T_{\rm{m, 0}}$ & Initial temperature at the top of the convecting mantle & $1500$ & $\unit{K}$ \\
$\Delta T_{\rm{cmb, 0}}$ & Initial temperature drop over the core-mantle boundary & $1500$ & $\unit{K}$  \\
$T_{\rm{s, 0}}$ & Surface temperature (constant) & $500$ & $\unit{K}$  \\
$D_{\rm{l, 0}}$ & Initial stagnant lid thickness & $50$ & $\unit{km}$  \\
$D_{\rm{l, min}}$ & Minimum stagnant lid thickness & $1$ & $\unit{km}$  \\
$\delta_{\rm{c, par}}$ & Lower thermal boundary layer thickness for overheated mantle & $10$ & $\unit{km}$ \\
$k_{\rm{m}}$ & Thermal conductivity of the mantle & $1$ & $\unit{W\, m^{-1} K^{-1}}$  \\
$k_{\rm{l}}$ & Thermal conductivity of the lithosphere & $1$ & $\unit{W\, m^{-1} K^{-1}}$  \\
$\alpha_{\rm{m}}$ & Thermal expansivity of the mantle & $2\times 10^{-5}$ & $\unit{K^{-1}}$  \\
$c_{\rm{m}}$ & Specific heat capacity of the mantle & $1200$ & $\unit{J\, K^{-1} kg^{-1}}$  \\
$c_{\rm{c}}$ & Specific heat capacity of the core & $800$ & $\unit{J\, K^{-1} kg^{-1}}$  \\
$A$ & Activation energy & $10^5$ & $\unit{J \, mol^{-1}}$  \\
$\eta_{\rm{min}}$ & Minimum mantle viscosity due to melting & $0.1$ & $\unit{Pa \; s}$  \\
$\mu_{\rm{max}}$ & Maximum mantle rigidity & $2\times10^{11}$ & $\unit{Pa}$  \\
$\mu_{\rm{min}}$ & Minimum mantle rigidity due to melting & $10^{-7}$ & $\unit{Pa}$  \\
$\Delta_{\eta}$ & Disaggregation width for viscosity & $0.01$ & --- \\
$\Delta_{\mu}$ & Disaggregation width for rigidity & $0.01$ & ---  \\
$\Phi_{\rm{D}}$ & Disaggregation point & $0.4$ & ---  \\
\hline
\end{tabular}
\end{center}
\end{table}

\subsection{Evolutionary paths of Proxima Centauri b}

\begin{figure}[ht]
  \begin{center}
    \includegraphics[width=1.\textwidth]{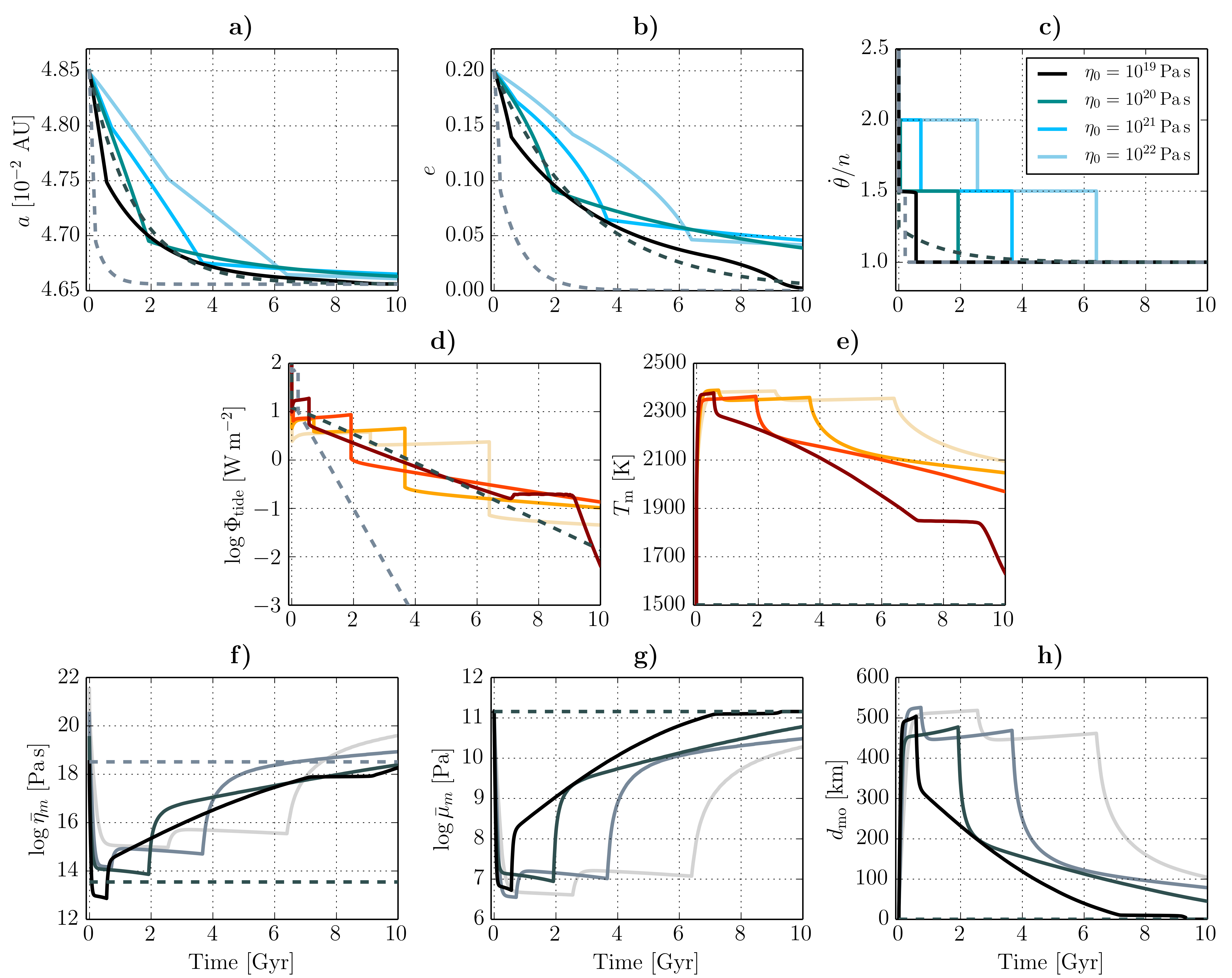}
  \end{center}
\caption{Simultaneous evolution of the spin-orbital parameters (\textit{upper row}), thermal state (\textit{middle row}) and interior properties (\textit{lower row}) of Proxima Centauri b. Going from the upper left corner to the lower right corner, the individual panels depict a) the semi-major axis, b) the orbital eccentricity, c) the spin-orbit ratio, d) the surface tidal heat flux, e) the mantle temperature measured under the stagnant lid, f) the average mantle viscosity, g) the average mantle rigidity, and h) the thickness of the magma ocean. Initial eccentricity was set to $e_0=0.2$ and reference viscosity spans from $\eta_0=\unit[10^{19}]{Pa\; s}$ (darkest colors) to $\eta_0=\unit[10^{22}]{Pa\; s}$ (lightest colors). For comparison, dashed lines indicate the spin-orbital evolution and the surface tidal heat flux for a model with constant (nonevolving) temperature profile. Dark gray dashed line corresponds to reference viscosity $\eta_0=\unit[10^{14}]{Pa\; s}$ and light gray dashed line to reference viscosity $\eta_0=\unit[10^{19}]{Pa\; s}$.} \label{fig:temps_proxima}
\end{figure}

Figure \ref{fig:temps_proxima} shows the coupled thermal-orbital evolution in the model case of Proxima Centauri b. For illustration purposes, we assume that the planet begins on a mildly eccentric orbit ($e=0.2$) and its initial semi-major axis is set to its presently observed value\footnote{The present-day semi-major axis $a_{\rm{now}}$ and the present-day eccentricity $e_{\rm{now}}$ are related to the initial values by $a_{\rm{now}}(1-e_{\rm{now}}^2)=a_{0}(1-e_{0}^2)$. Nevertheless, since the present-day eccentricity of the exoplanets, as well as the age of their host stars, are known with relatively large errors, we cannot exactly trace back the $a_{0}$ corresponding to the chosen values of $e_{0}$. Although it would be possible to test different initial semi-major axes, we decided to set $a_{0}$ for all model cases of a given system to the same value and to vary only the initial eccentricity.}. To include the unknown effect of pressure on the lower mantle viscosity, we consider four possible reference viscosities $\eta_0$ in the range from $10^{19}$ to $\unit[10^{22}]{Pa\; s}$. In addition to mimicking different pressure dependencies, the range of viscosities also accounts for different possible mineralogical compositions of the mantle. To identify the effect of the evolving interior, we further run two additional simulations, in which the interior temperature profile and the rheological properties are held constant while the orbital parameters evolve (dashed lines in Figure \ref{fig:temps_proxima}). The comparison between the coupled model and the constant-interior model is discussed at the end of this subsection.

In the beginning, the planet despins rapidly into the first stable spin-orbit resonance. Depending on the reference viscosity, it ends up either in the $2:1$ resonance ($\eta_0=10^{21}$ or $\unit[10^{22}]{Pa\; s}$) or in the Mercury-like $3:2$ resonance ($\eta_0=10^{19}$ or $\unit[10^{20}]{Pa\; s}$). The despinning phase is, furthermore, marked by a rapid increase in the interior temperature---a consequence of relatively high orbital eccentricity and nonsynchronous rotation. Due to its orbital configuration, the planet undergoes considerable tidal loading and the dissipated heat remains in the mantle, since it cannot be efficiently taken away by the convection. This period of overheating is, however, only transient. As can be seen in the lower row of Figure \ref{fig:temps_proxima}, the increase in the interior temperature is accompanied by a similarly steep decrease in the average mantle viscosity and rigidity. We note that in these model settings, the top of the mantle melts shortly after the beginning; the melting temperature at the relevant pressures is around $T_{\rm{m}}=\unit[1900]{K}$. The melt is predominantly concentrated below the lid as both solidus and liquidus temperature increase considerably with depth.

The increase in the upper mantle temperature and decrease in the average viscosity and rigidity continues until the mantle reaches an equilibrium state. Once the viscosity decreases to a value that enables efficient transport of the generated heat to the surface, the interior temperature stops rising and---independently of the reference viscosity---stabilizes around $\unit[2350]{K}$. Increased mantle temperatures also affect the efficiency of heat transfer from the core. As the mantle reaches thermal equilibrium at high temperatures, such as in this model case, the cooling rate of the core substantially decreases and the core attains a quasi-equilibrium. At this moment, the planet possesses a $400-\unit[500]{km}$ thick magma ocean and the global melt fraction in the mantle is about $25~\%$. How long the interior remains in the equilibrium state depends on the evolution of spin-orbital parameters. Decreased mantle viscosity and evolving eccentricity act together in destabilizing the actual spin-orbit resonance. As was already illustrated in the previous subsections, the regions of stability for distinct stable spin states depend on the average mantle viscosity (or rigidity) and on the eccentricity. Different model cases in Figure \ref{fig:temps_proxima} thus undergo the transition to lower spin-orbit resonances at different times.

Each transition between stable spin-orbit resonances is accompanied by a drop in the tidal heat rate. Decreased heat rate results in slower orbital evolution (noticeable as a change of the slope in the first two panels) and in fast cooling of the vigorously convecting mantle. The subsequent fate of the planet's interior depends on the new spin state. If the planet despinned into yet another nonsynchronous spin-orbit resonance, such as $3:2$ for the higher considered viscosities, the tidal dissipation remains an important, almost eccentricity-independent heat source (see Figure \ref{fig:eta_ecc}). In this case, the interior promptly acquires a new equilibrium thermal state at slightly lower temperature and higher average viscosity. On the other hand, if the planet despins directly into the synchronous rotation, the tidal heat rate becomes strongly dependent on the eccentricity. The interior first cools down to a quasi-equilibrium state, in which the mantle temperature continues to decrease. The long-term cooling is then controlled by the decaying eccentricity.

A noteworthy feature of the depicted evolutionary paths is the late thermal equilibrium of the model case with $\eta_0 = \unit[10^{19}]{Pa\; s}$. The sudden increase in the surface tidal heat flux, which occurs after $\unit[7]{Gyr}$, is associated with partial crystallization of the remnant magma ocean. Once the magma layer begins to crystallize and the local rigidity in the upper mantle increases, it becomes a significant source of tidal dissipation, able to counterbalance the gradual cooling. The period of thermal equilibrium is, however, terminated after $\unit[2]{Gyr}$. By the end of this transient phase, the average melt fraction in the magma layer decreases to $0.4$ and the rigidity slowly increases. At $\unit[9.2]{Gyr}$, the ocean eventually disappears. After leaving the equilibrium, the planet follows the path of gradual cooling down with even steeper slope.\\

The presented model of Proxima Centauri b illustrates, in the first place, the principal role of tidal locking in the long-term thermal and orbital evolution. Despinning into a new spin-orbit resonance affects all other studied quantities and enables abrupt changes in the slope of the semi-major axis and the eccentricity. The eccentricity, in turn, complements the effect of the planet's rotation in determining the tidal heat rate.

The combined effect of the two parameters is most prominent in the evolution of the average surface tidal heat flux (panel "d" in Figure \ref{fig:temps_proxima}). In the beginning, the highest rate of tidal heating is observed in the model case with the lowest reference viscosity. The highest viscosity case, on the other hand, dissipates the lowest amount of energy. Since the eccentricity of both cases is comparable and the rotation is nonsynchronous, the difference lies in the different susceptibility to tidal deformation. After $\unit[5]{Gyr}$, however, the situation almost reverses. While the highest reference viscosity case remains in the $3:2$ resonance for a considerable time and supports persistent tidal heating, the lowest reference viscosity case has already despinned into the synchronous rotation and dissipates order of magnitude less energy. The former case also retains more than two times higher orbital eccentricity than the latter, which further contributes to the increased tidal dissipation.

Another interesting observation can be made by comparing the orbital eccentricities over the last $\unit[4]{Gyr}$. Independently of the interior properties, the three highest-viscosity cases end up on very similar orbits. The same tendency was also observed in additional model runs with different initial eccentricities ($e=0.05$, $0.1$ or $0.4$). The resulting eccentricities after $\unit[10]{Gyr}$ of evolution tend either to a similar nonzero value (between $0.03$ and $0.05$) or---in the lowest reference viscosity cases---towards circular orbit. However, as was illustrated in this subsection, the earlier evolution of all model parameters is relatively complex and cannot be described by a simple rule.\\

Compared to the model with fixed interior properties (dashed lines in Figure \ref{fig:temps_proxima}), the coupled model generally maintains higher orbital eccentricities. To illustrate this, we first focus on the evolution of the "fixed-interior" model case with reference viscosity $\eta_0 = \unit[10^{19}]{Pa\; s}$. In the beginning, the chosen model case is confronted with much higher tidal dissipation than any of the other model cases. Since it is not moderated by a decrease in the rheological parameters, this overheating results in rapid circularization of the orbit and tidal locking into synchronous rotation within the first billion of years. The surface tidal heat of the chosen model case then quickly becomes negligible.

The orbital evolution of the second fixed-interior case, with $\eta_0 = \unit[10^{14}]{Pa\; s}$, resembles the pattern of the evolving-interior model with $\eta_0 = \unit[10^{19}]{Pa\; s}$. However, while the latter possesses a relatively stiff mantle, whose average viscosity is reduced only by the presence of a magma ocean, the former has low mantle viscosity by definition. Hence, while the evolving-interior model gets early locked into the 1:1 resonance, which reduces its rate of orbital evolution, the fixed-interior model remains in stable pseudosynchronous rotation. Although the two model cases in question provide similar results for both of the studied orbital parameters, the difference in the rotation history would yield substantially different atmospheric forcing and different surface conditions. This example illustrates that, when studying the spin rate evolution of partially molten bodies, the assumption of homogeneous vs layered mantle may lead to qualitatively different results.

\subsection{Thermal and orbital state of evolved low-mass exoplanets} \label{ssub:r3all}

The long-term evolution of all chosen exoplanets follows similar tendencies. Depending on the initial orbital eccentricity, they experience one or more spin-orbit lockings and possibly also a serie of thermal equilibria. In the previous subsection, we inspected the evolutionary path of Proxima Centauri b over $\unit[10]{Gyr}$. The actual age of the system is, however, much lower. According to asteroseismic observations of the $\alpha$ Centauri binary \citep{thevenin02}, it originated $\unit[4.85]{Gyr}$ ago and, therefore, is only a few hundred million years older than the Sun. Since the age of the other two exoplanets is---to the best of our knowledge---currently unknown, we now focus on the evolutionary outcome of several model cases after $\unit[5]{Gyr}$.

\begin{figure}[ht]
  \begin{center}
    \includegraphics[width=1.\textwidth]{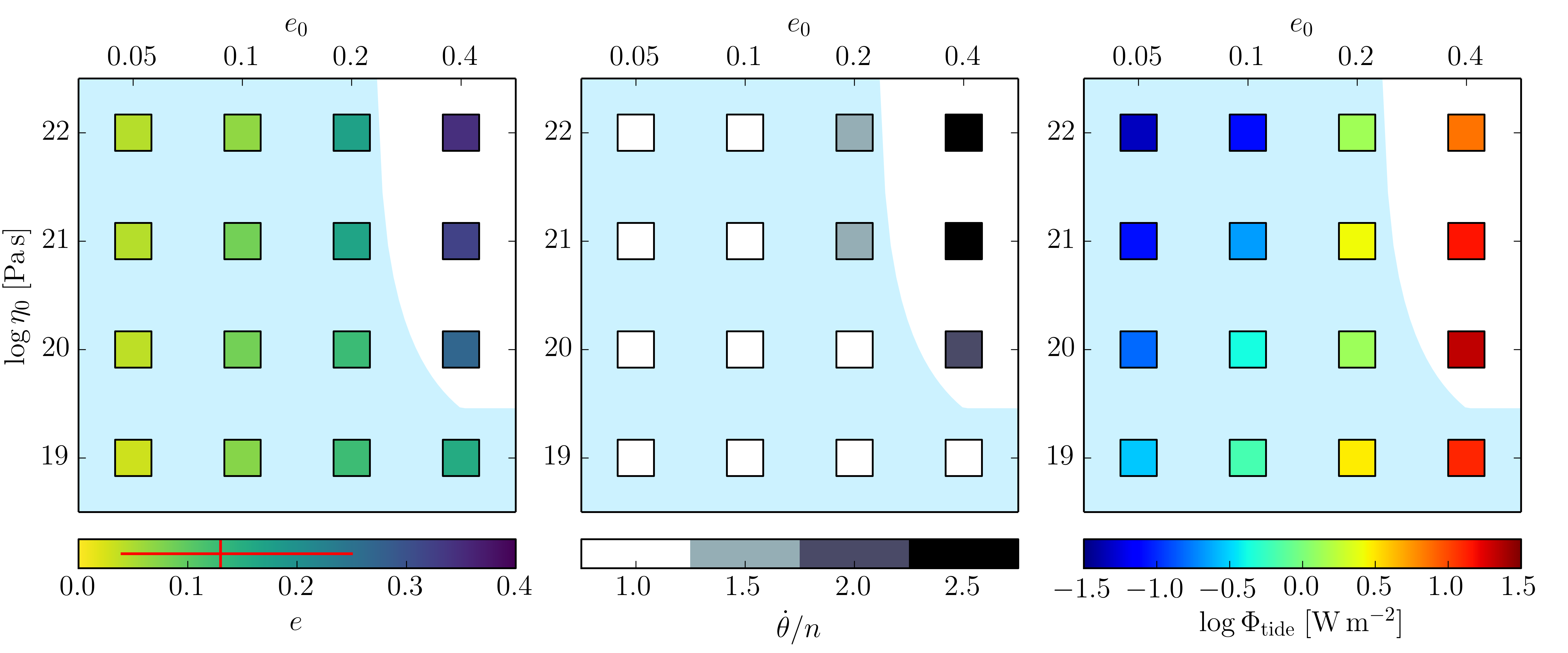}
  \end{center}
\caption{Orbital and thermal characteristics of GJ 625 b after $\unit[5]{Gyr}$ of evolution. Depending on the initial orbital eccentricity $e_0$ ($x$-axes) and the reference mantle viscosity $\eta_0$ ($y$-axes), the individual panels illustrate the evolved eccentricity (\textit{left}), the spin-orbit ratio (\textit{middle}) and the surface tidal heat flux (\textit{right}). Light blue areas correspond to the model parameters for which the evolved eccentricity complies with observation \citep{suarez17}. The range of the empirically given values is also indicated by a red line in the first colorbar.} \label{fig:gj625_evol}
\end{figure}

\begin{figure}[ht]
  \begin{center}
    \includegraphics[width=1.\textwidth]{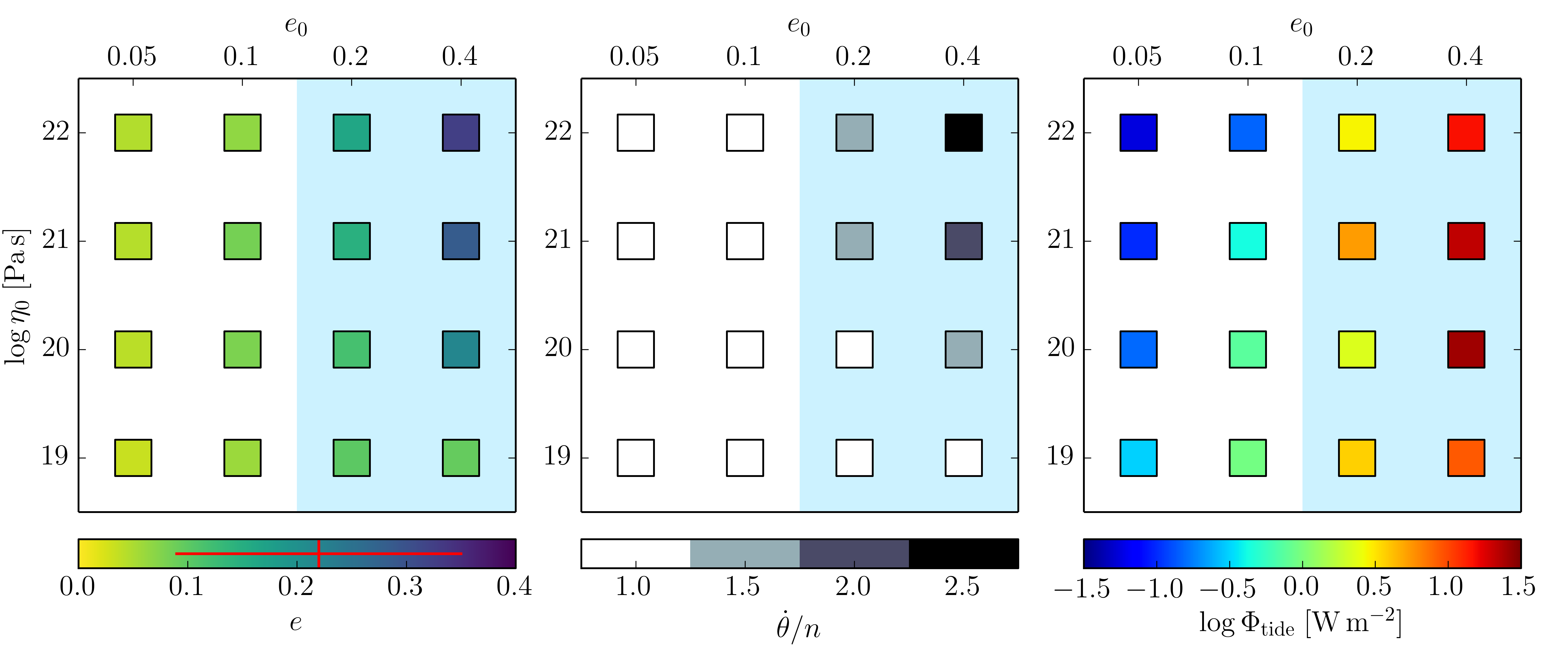}
  \end{center}
\caption{Same as Figure \ref{fig:gj625_evol}, but for GJ 411 b. The span of empirically given eccentricities is taken from \citet{diaz19}.} \label{fig:gj411_evol}
\end{figure}

\begin{figure}[ht]
  \begin{center}
    \includegraphics[width=1.\textwidth]{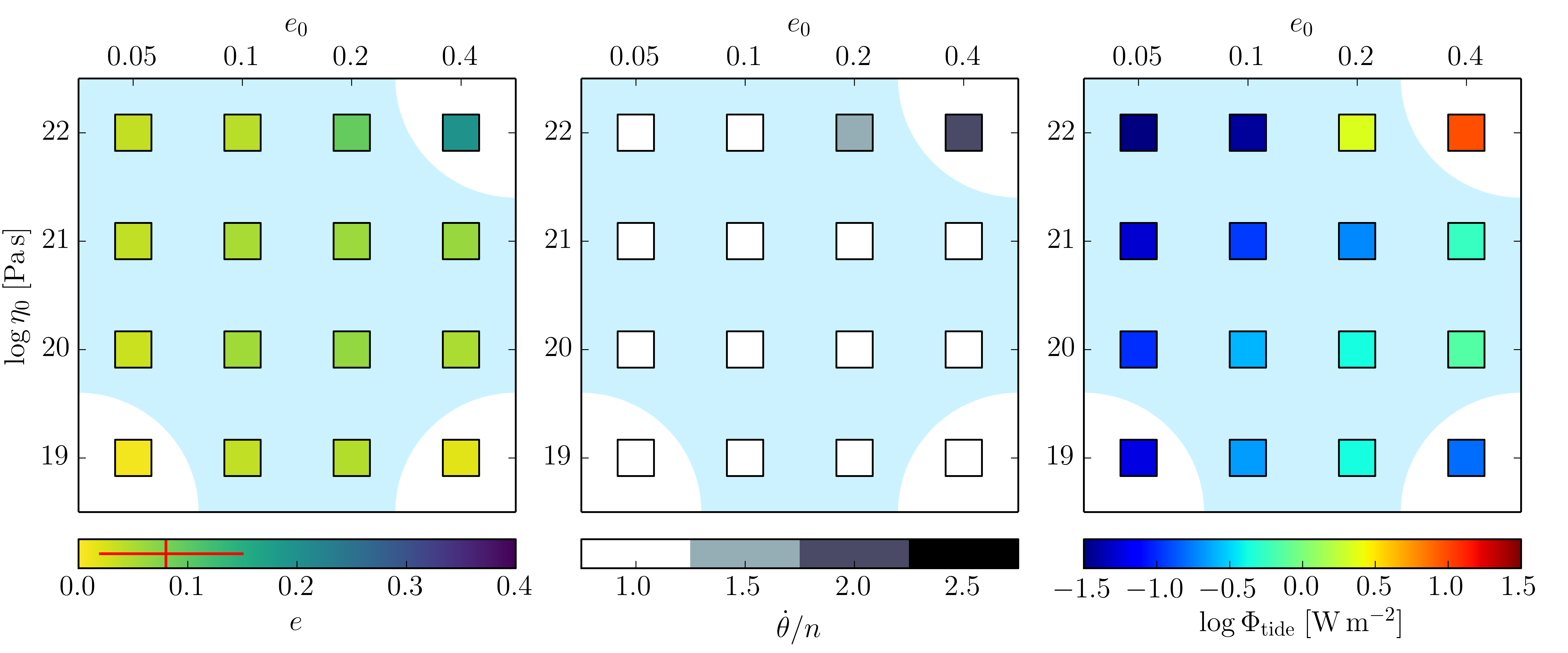}
  \end{center}
\caption{Same as Figure \ref{fig:gj625_evol}, but for Proxima Centauri b. The span of epirically given eccentricities is taken from \citet{jenkins19}.} \label{fig:proxima_evol}
\end{figure}

For each of the three studied exoplanets, we consider four possible initial eccentricities ($e_0=0.05$, $0.1$, $0.2$ and $0.4$) and four reference viscosities identical to the previous subsection. The tidally evolved eccentricity after $\unit[5]{Gyr}$, together with the spin-orbit ratio and the surface tidal heat flux, is depicted in Figures \ref{fig:gj625_evol} to \ref{fig:proxima_evol}. To sort out the model outcomes which do not comply with the eccentricity derived from current observations, we indicate the plausible eccentricities by a red line in the colorbar and by light blue background on the individual panels.\\

Figure \ref{fig:gj625_evol} illustrates the possible evolution outcomes of GJ 625 b: a planet that lies on the inner edge of the habitable zone. Its instantaneous, tidally evolved eccentricity follows a predictable pattern, which is only mildly affected by the reference viscosity. The higher the initial eccentricity, the higher its value after $\unit[5]{Gyr}$. Since the uncertainty of the empirically determined eccentricity is relatively high, the majority of the models comply with the observation. Model cases initialized to the highest considered eccentricity, possibly due to external excitation, are, however, excluded. The second panel of the figure indicates that the planet is most probably locked in the synchronous rotation. Only in the rare case of high-eccentricity start with Earth-like mantle viscosity does the planet sustain the $3:2$ resonance. A consequence of the resonance locking is further reflected in the nontrivial dependence of the surface tidal heat flux on the reference viscosity. While in the lower eccentricity cases the resulting heat flux monotonically increases with decreasing viscosity, for $e_0=0.2$ and $e_0=0.4$ the rheological parameters play a lesser role than the rotation state. The values of the tidal heat flux also indicate that the evolved planet is most probably less volcanically active than Io. If rotating synchronously, its thermal output due to tides might be hypothetically comparable with other internal sources of the heat (e.g., radiogenic heating or remnant heat from accretion).

Due to the similar orbital periods and masses, the conclusions given for the model of GJ 625 b are applicable also to the case of GJ 411 b (Figure \ref{fig:gj411_evol}). The difference, however, lies in the eccentricities. For GJ 411 b, the mean value of the empirically given eccentricity is higher than $0.2$, which points at currently high surface tidal heat flux and nonsynchronous rotation. Were the present day orbit influenced only by tides, the planet would have to originate on a highly eccentric trajectory. The evolved spin rate predicted by the coupled model ranges from the $1:1$ spin-orbit resonance for low reference viscosities up to the $5:2$ resonance for higher viscosity values. Accordingly, the surface tidal heat flux is expected to surpass the activity of Io.

The orbital and physical parameters of the Proxima Centauri system are rather unlike the two previously described exoplanets. First of all, its lower mass and lower predicted orbital eccentricity make it akin to the Earth. Figure \ref{fig:proxima_evol} suggests that independently of the initial eccentricity and the reference viscosity, the planet's orbit tends to a relatively low eccentricity below $e=0.1$. Hence, the majority of the cases comply with the current observations. As a consequence of the low resulting eccentricity, the model predicts the prevalence of synchronous rotation. The only exception from this pattern is the $3:2$ resonance expected for a high reference viscosity model with initial eccentricity of $e_0=0.2$ (see also Figure \ref{fig:temps_proxima}). As most of the model cases end up in the same rotation state, the thermal output of Proxima Centauri b is usually determined by the actual eccentricity and the rheological parameters. Especially for low initial eccentricities, the surface tidal heat flux is comparable to the total heat flux on the Earth and, depending on the other heat sources and on the effect of the subsurface magma layer, it might not present an obstacle for the hypothetical habitability.

In addition to the presented diagrams, it is worth noting that the majority of the model cases support a long-lived magma ocean. The only case in which the ocean disappears during the first $\unit[5]{Gyr}$ is Proxima Centauri b with the lowest considered initial eccentricity and reference viscosity. However, since the effect of the subsurface magma ocean on the surface conditions is beyond the scope of this paper, we postpone the discussion of this phenomenon to a more detailed study of the interior evolution.
 
\section{Discussion} \label{sec:disc}

The spin-orbital dynamics of a tidally loaded exoplanet interact with its interior evolution in an intricate way. Throughout the previous section, we attempted to illustrate the complexity of the coupled model, which is given primarily by the viscoelastic rheology and the evolution of the interior structure. Although the complexity is inherent to the nature of the problem, the exact results and predictions depend on the assumptions made. In this section, we focus on the main features of the model which might have affected its outcome, and we also discuss the implications of our results.

\subsection{Stable spin states}

The thermal and orbital evolution of viscoelastic planets on eccentric orbits is interrelated with the evolution of their spin state. As a consequence of the viscoelastic behaviour, the planets with rocky interiors tend to the proximity of spin-orbit resonances, whose stability is given by the frequency of tidal loading and by the rheological parameters \citep[e.g.,][]{ferrazmello13,makarov13,correia14,ferrazmello15}. Different orbital configurations (e.g., the eccentricity) yield a different spectrum of loading frequencies and determine the actual stable spin state. Although the basic aspects of the coupled system's dynamics can be captured by models with synchronous rotation \citep[e.g.,][]{henning09}, the results of Sections \ref{sec:results1} to \ref{sec:results3} show that the consideration of higher spin-orbit resonances is important, especially in the case of planets with low orbital eccentricity. In this case, the eccentricity tides are already weak and the tidal dissipation is sustained primarily by the nonsynchronous rotation. The time at which the planet undergoes a transition to the synchronous rotation determines the values of the terminal, slowly evolving orbital parameters.

The stable spin state of real moons and close-in planets is given not only by the gravitational tides but also by the thermal atmospheric tides \citep[e.g.,][]{gold69,auclair19} and/or by their permanent deformation \citep[e.g.,][]{goldreich66}. In the presence of a significant atmosphere, as is the case for the planet Venus, the dayside experiences higher temperatures and lower atmospheric pressures than the nightside. As a consequence of the periodic thermal forcing and the redistribution of atmospheric masses, the planet becomes subjected to additional tidal torque acting on its atmosphere. The competition between the gravitational and thermal tidal torques destabilizes spin-orbit resonances and drives the planet to nonsynchronous rotation \citep[e.g.,][]{leconte15}. Similarly, in the case of a triaxial planet, the tidal torque is counterbalanced by a torque acting on the permanent deformation. The inherent triaxiality may either stabilize the planet in an otherwise unstable synchronous rotation \citep[e.g.,][]{goldreich66,goldpeal66} or, conversely, it may prevent the planet from exact synchronization by locking it into a higher spin-orbit resonance \citep[e.g.,][]{makarov12,zanazzi17}.

When calculating the stable spin state, we also assumed that the spin axis is perpendicular to the orbital plane. The tidal effects in our study were, therefore, only due to the nonzero eccentricity. In general, the rotation state is determined by the planet's figure and interior structure and its equilibrium obliquity corresponds to one of the Cassini states \citep[e.g.,][]{peale69,boue20}. The stability of individual Cassini states also depends on the configuration of the planetary system. A planet with nonzero obliquity might be attracted to different spin-orbit resonances \citep{boue16}, and its thermal budget is then enhanced by obliquity tides \citep[e.g.,][]{peale78}. Both of these effects contribute to long-term evolution. As a first-order approximation, the ratio of the obliquity heating to the eccentricity heating can be expressed as \citep[e.g.,][]{peale78,chyba89,murray99}

\footnotesize
\begin{equation}
\frac{P^{\rm{tide}}_{\beta}}{P^{\rm{tide}}_{e}}\, \simeq\, \frac{\sin^2\beta}{7e^2}\; .
\end{equation}\normalsize\\
Stable nonzero obliquity may, therefore, prevent runaway cooling once the orbital eccentricity decreases to a negligible value. A test calculation with both sources of tidal heating and a constant obliquity of $20^{\degr}$ indicates that increased tidal heat rate stabilizes the interior at a higher temperature, with a thicker magma ocean. The orbital evolution is marked both by the increased dissipation and by the earlier destabilization of higher spin-orbit resonances due to low mantle viscosity. However, in the long term, the obliquity of strongly tidally loaded exoplanets \citep[e.g.,][]{fabrycky07} tends to zero.

\subsection{Sources of orbital eccentricity}

The orbital eccentricities of moons and planets in the Solar System are shaped mainly by mutual interactions between the bodies. A well-known example of this effect is planet Mercury, whose eccentricity may rise up to $1$ due to the gravitational action of other planets \citep[e.g.,][]{laskar94,batygin08,laskar09,lithwick11,boue12}, and large satellites of gas giants, with orbital parameters forced by orbital resonances \citep[e.g.,][]{schubert10}. A considerable change of the satellites' orbits might have also been caused by transient events, such as close encounters of giant planets during the period of planet migration \citep[e.g.,][]{deienno14}. In analogy with the Solar System, the detection of nonzero orbital eccentricities among close-in exoplanets is often a consequence of ongoing gravitational perturbations by other bodies \citep[e.g.,][]{takeda05,pu18,vaneylen19} or a relic of past catastrophic events, such as planet-planet scattering \citep[e.g.,][]{petrovich14,huang17}. In single-planetary systems or systems without substantial gravitational forcing, the orbital eccentricity can be explained by initial conditions during planet formation and by subsequent evolution in a gas disk \citep[e.g.,][]{kley12,ragusa18}, as well as by formation in an unstable multiplanetary system. Another possible source of nonzero orbital eccentricity is tidal interaction of a close-in planet with a rapidly rotating host star \citep[][equation (156)]{boue19}. For a detailed overview of the eccentricity excitation mechanisms, we refer the reader to, e.g., \citet{namouni07}.

In this paper, we assumed that the planet begins on an eccentric orbit and evolves only under the action of tides. The three studied exoplanets, GJ 625 b, GJ 411 b and Proxima Centauri b, were also chosen on the grounds of absent, negligible or yet unknown gravitational forcing by a third body. Applying the tidal model to such exoplanets may help to shed light on the initial conditions in the system or constrain previous gravitational forcing the system underwent. The results of the parametric study presented in Subsection \ref{ssub:r3all} indicate that the empirically given eccentricities can be reconciled with a wide range of initial eccentricities and reference viscosities. In the case of exoplanets GJ 625 b and GJ 411 b, this is mainly due to the large errorbars of the current eccentricities. For Proxima Centauri b, the reason also lies in the similar tendencies of the test cases, all of which evolve towards a similar, mild eccentricity after $\unit[5]{Gyr}$.

The most specific of the studied exoplanets is GJ 411 b. Its high present-day eccentricity determines it either to a high-eccentricity start or to eccentricity excitation in the past. An~additional constraint on the initial conditions would be given by the age of the host star, which is currently unknown. Although GJ 411 b, discovered in 2019, is the only confirmed exoplanet in the system, the star GJ 411 had been for a long time suspected to host a planetary companion. However, none of the previously reported detections has been confirmed and the existence of other bodies in the system is putative \citep[see also discussion and references in][]{diaz19}. In the presence of strong eccentricity forcing by a second planet, the thermal evolution of GJ 411 b would be affected by periods of high tidal dissipation and its evolution would resemble that of resonant moons in the Solar System.

\subsection{Parametrized convection and melting model}

In Section 8, we observed that the coupled thermal-orbital evolution proceeds as a sequence of thermal equilibria. In each equilibrium, the heat sources due to the tidal dissipation are entirely compensated by the heat loss by mantle convection. However, the heat loss is determined by the selected convection regime. A different cooling rate would yield different equilibrium states, different temperature profiles and, most importantly, different rheological properties. Here, we shall discuss our choice of the convection regime and its impact on the resulting cooling rates.

First of all, we note that the choice of the convection regime is a very complex problem that should be ideally addressed by numerical modeling. The thermal or thermo-chemical convection in the Earth and other Solar System bodies is discussed in an extensive literature. Despite the scarcity of information on exoplanets, numerical models are used to investigate the scaling of mantle convection with mass and radius (including the effect of extreme pressure) and to describe the convection regimes in massive terrestrial exoplanets \citep["super-Earths"; e.g.,][]{vandenberg10,cizkova17}. The probability of plate tectonics on super-Earths was investigated by, e.g., \citet{vanheck11}, \citet{foley12} and \citet{noack14}. \Citet{summeren11} studied the mantle convection in tidally locked terrestrial planets with large surface temperature contrasts. The presence of huge volumetric heating or hot conditions, as experienced by close-in planets, brings additional challenges due to large-scale melting and the emergence of magmatic ponds and oceans. \citet{vilella17} used a systematic approach to build a diagram providing conditions for partial melting based on the planet size and internal heating. An ideal solution accounting for substantial internal heating or extreme temperatures on close-in exoplanets is a 3d multiphase convection with tidal dissipation as a source of volumetric energy. Nevertheless, the inclusion of the melt-solid phase interaction, such as the melt migration, melt production and recrystallization, requires complex description \citep[e.g.,][]{bercovici01} and possibly leads to extremely computational demanding simulations.

Any 3d/2d modeling efforts are thus beyond the scope of this study. In order to understand the main aspects of secular thermal-orbital coupling, we follow here a traditional approach of parametrized convection, where we have to account for possibly significant melting. Depending on the size of the planet and the magnitude of internal heating, the incorporation of melt into the model can be treated in different ways. Parametric studies of mantle convection in Mars or stagnant-lid Earth \citep{breuer06,tosi17}, on which we base our interior evolution model, assume that the melt with positive buoyancy is instantaneously extracted from the mantle and becomes a building material for the crust. This approach might result in depletion of the mantle material and in its dehydration \citep[e.g.,][]{plesa12}, which increases the upper mantle solidus and regulates further production of the melt. Coupled thermal-orbital models focused on small terrestrial exoplanets \citep{henning09,shoji14,driscoll15}, on the other hand, assume well-mixed mantle with evenly distributed melt and decreased viscosity and rigidity of the entire planet. However, the neglection of radial stratification of the planet might substantially affect the resulting tidal heating pattern \citep{henning14}.

A realistic parametrization of the subsurface melt dynamics would take into account the permeability of the lithosphere \citep[e.g.,][and references therein]{spiegelman93} and the melt buoyancy, which is a complex function of the mantle composition. According to experimental studies with floating olivine in silicate liquids, the melt becomes neutrally buoyant around $\unit[7-12]{GPa}$, at the density crossover of the two phases \citep[e.g.,][]{agee93,ohtani95,agee08}. Specifically, in the upper mantle of the Earth, the existence of a density crossover might enable the formation and maintenance of hydrous melts above the $\unit[410]{km}$ discontinuity \citep{agee08}. In a general case, the position of this transition depends on the water content, the mineralogy of the mantle and the local temperature. Since all of these parameters vary with depth, the realistic incorporation of melt migration would require a much more detailed study. Another important mechanism, which affects the cooling rate of the planet, is global-scale volcanism \citep[e.g.,][]{oneill07}. As proposed by \citep{moore17}, moons and planets with overheated and partially molten mantles might be effectively cooling down by heat-pipe volcanism, before they transition to the stagnant lid or mobile lid regime. Strongly tidally loaded bodies, such as Jupiter's moon Io \citep{moore03}, or large terrestrial exoplanets \citep{moore17} might remain in the stage of heat pipe for billions of years.

We opted here for parametrized stagnant lid convection with a very simplistic treatment of the melt. The model does not consider any melt migration and, conversely, assumes a stable magma layer in the range of depths where the local temperature exceeds the disaggregation point. This assumption enables us to assess the dynamical effect of a liquid, almost nondissipative layer above much more viscous lower mantle. The liquid layer also decouples the lithosphere and the rest of the mantle, which is then more susceptible to tidal deformations. Nevertheless, while we consider the magma ocean in the tidal model, it is included only in a simplified manner in the mantle convection. The maintenance of the magma ocean can be understood as a limit case for interior evolution, maximizing the effect of partial melting. The presence of melt decreases the upper boundary layer thickness via the geometric average of the mantle viscosity and allows for large heat flux into the lithosphere. This results in faster cooling of the mantle, as expected for any presence of melt. The partial melting and the formation of magma ocean also help to regulate the thermal runaways by the following two mechanisms: i) the consumption of a part of the tidal heat by the phase transitions and ii) the change of the rheological properties. During the cooling part of the evolution, the magma ocean can delay the cooling due to recrystallization and latent heat.

Finally, we should note that our model neglects any circulation in the magma ocean. The tidal response of fluid is essentially different from the response of solid layers and should be calculated by a different set of tidal equations \citep[such as Laplace tidal equations; see, e.g.,][]{tyler15}. Dissipation in the liquid layers is directly affected by the rotation rate and is characterized by the formation of inertial waves \citep[e.g.,][]{roviranavarro19}. Several recent studies have investigated the tidal response of realistic liquid layers in Io or icy satellites of Solar System gas giants. The importance of tidal dissipation in subsurface oceans generally depends on the thickness of the ocean and the thickness of the overlying shell, which tends to dampen the ocean tides \citep{beuthe16,matsuyama18}. Additional heat can be also produced by turbulent dissipation, internal gravity waves in the ocean and interaction of the fluid with the ocean basin topography.

\subsection{Habitability of tidally evolving exoplanets}

Two of the planets chosen for our parametric study, namely GJ 625 b and Proxima Centauri b, are reported to reside in the conventional habitable zone of their host star \citep{suarez17,anglada-escude16}. The habitable zone is conventionally defined as the range of orbital distances that allow the planet to sustain liquid surface water under certain atmospheric conditions \citep{kasting93}. The boundaries of a habitable zone are given by the incident flux and thus depend on the stellar type. Later refinements of the original definition relate the boundaries of the habitable zone to additional parameters, such as the planetary mass, atmosphere and orbital eccentricity \citep[e.g.,][]{seager13,kopparapu13,kopparapu14,palubski20}, or restrict the range of plausible temperatures to allow for the formation of complex organics \citep{wandel18,wandel20} or even complex life \citep{schwieterman19}. Although the orbit's location inside the habitable zone may serve as an initial guess on the surface conditions of an exoplanet, it is not sufficient to determine its potential to harbour life. Planetary habitability is influenced by a combination of many effects, some of which are also altered by tides \citep[e.g.,][]{seager13,kane17,lingam18,delgenio19}.

Tidal evolution affects habitability in both positive and negative ways. Secular shrinking and circularization of the orbit may drive the planet inside or outside the habitable zone \citep{barnes09,palubski20} and the tidal alignment of the spin axis may principally influence its climate \citep{heller11}. Tidal strain in the lithosphere can be vital in developing plate tectonics on close-in worlds \citep{zanazzi19}, while strong tidal dissipation is able to transform the planet into an inferno \citep{barnes13}. An essential question determining the habitability of tidally evolving exoplanets is the effect of spin-orbit coupling on the planetary climate and surface conditions. Synchronous rotation, which is the most probable spin state of planets with low orbital eccentricities, yields extreme differences in the insolation of the surface and in the surface temperatures \citep{dobro07}. Uneven heating of the atmosphere, combined with active volcanism, may trigger various feedbacks able to destabilize the climate \citep[e.g.,][]{kite11}. However, it may also prevent the planet from going through periods of global glaciation \citep{checlair19}. 

Another tidal phenomenon able to affect planetary habitability is the overheating and melting of the interior. Partial melting and the subsequent volcanic outgassing can gradually enrich the planetary atmosphere in greenhouse gases, such as CO$_2$ \citep{dorn18}, which may be vital for planets whose atmospheres were eroded during the early active phase of stellar evolution \citep[e.g.,][]{loyd18}. Although secular partial melting results also from radiogenic heating, tidal dissipation may be an important additional source, especially in the mantles of higher-mass terrestrial planets (>$3\; M_{\earth}$), whose solidus temperatures are increased due to higher subsurface pressures \citep{noack17,dorn18}. Outgassing and melt extraction play an important role also in the recycling of planetary material. The carbon-silicate cycle, which contributes to the long-term climate stability of the Earth, is enabled by the interplay between active volcanism and temperature-dependent weathering \citep{walker81}. Carbon, outgassed to the atmosphere by volcanism, is later deposited into the crust and returned to the mantle by subduction. On rocky exoplanets in the stagnant-lid regime, recycling may be limited by the absence of an efficient mechanism drawing carbon down from the atmosphere. Nevertheless, the maintenance of the carbon-silicate cycle on such planets is not completely excluded \citep{foley18,valencia18}.

In this study, we assumed that the model planets operate in the stagnant lid regime and that the melt is not extracted by volcanism. While the conclusions of the coupled thermal-orbital model are marked by these assumptions and their answer on the question of planetary habitability might be limited, the results of our parametric study with fixed parameters gives us some insight into the tidal effect on the surface conditions of GJ 625 b and Proxima Centauri b. If the melt was extracted from the upper mantle and the model planets retained Earth-like rheological parameters (i.e., $\eta_{\rm{m}}\approx\unit[10^{21}]{Pa\;s}$ and $\mu_{\rm{m}}\approx\unit[10^{11}]{Pa}$), their surface would be potentially habitable only for low values of orbital eccentricity. On even slightly eccentric orbits, consistent with the observation, the surface tidal heat flux would exceed the values measured for Io. However, it should be noted that the orbital eccentricity of such strongly dissipating exoplanets would need to be maintained by external forcing, in order not to disappear during the first $\unit[1]{Gyr}$ of tidal evolution. In terms of the present-day surface tidal heat flux, the results of the coupled thermal-orbital model seem more optimistic than the results of the model with fixed parameters. For mild initial eccentricities, the surface tidal heat flux tends to values comparable with the total heat production of the Earth and tidal dissipation alone does not pose a serious obstacle for potential habitability.

\section{Conclusions} \label{sec:concl}

In this paper, we investigated the interconnection between the spin-orbital dynamics and thermal evolution of low-mass exoplanets around M-type stars. The planets were modeled as differentiated bodies with three or four homogeneous layers, whose mantle is described by the Andrade viscoelastic model. Consistently with the evolution of the mantle temperature profile, the model bodies were allowed to build up a stable subsurface magma ocean, which influenced the effectivity of heat transport as well as the tidal response. In addition to the coupled model, we also conducted several parametric studies with fixed interior structure (without the magma ocean) for an Earth analogue and for three low-mass exoplanet candidates. The purpose of these studies was to illustrate the effect of rheological, orbital and physical parameters of the planet on its evolutionary path, highest stable spin state, and on the surface tidal heat flux. The following summary highlights the main conclusions of this work:\\

1) The stability of individual spin-orbit resonances within the Andrade model is a complex function of the rheological parameters and the eccentricity. The resulting pattern (Figures \ref{fig:etamu_e005}-\ref{fig:eta_ecc}) depends on the interplay between the self-gravity and the rheological parameters and on the role of viscoelasticity at given tidal frequencies (see Appendix \ref{app2}). Especially at low orbital eccentricities, the despinning from higher spin-orbit resonance to the synchronous rotation results in a significant drop in the tidal heating. The secular tidal torque also depends on the planet's mass and radius. Planets with smaller radii and/or low core mass fractions ($\rm{CMF}$) tend to get locked into higher spin-orbit resonances than larger and/or more massive planets with the same rheological and material parameters of the interior layers.

2) For the range of eccentricities consistent with observation, the three studied low-mass exoplanets (GJ 625 b, GJ 411 b, and Proxima Centauri b) are able to maintain higher than synchronous spin-orbit resonances. Locking into such spin state would provide the planetary surface with relatively uniform insolation, which might have important consequences for the dynamics of the atmosphere and for the hypothetical habitability of the planet. However, as far as we consider a model with a homogeneous solid mantle (Section \ref{sec:results2}), this vital effect of the higher spin-orbit resonances may interfere with the increased tidal dissipation at even mild orbital eccentricities. For illustration, the model of Proxima Centauri b with a reference Earth-like rheology ($\eta_{\rm{m}}=\unit[10^{21}]{Pa\, s}$, $\mu_{\rm{m}}=\unit[200]{GPa}$, no magma ocean) might be locked into 3:2 resonance for $e=0.08$ and into 2:1 resonance for $e=0.15$. Nevertheless, in both cases, the surface tidal heat flux exceeds the values reached on Io. A more optimistic result is obtained either at the lowest considered orbital eccentricity $e=0.02$ (with 1:1 resonance) or with a partially molten interior.

3) In the coupled thermal-orbital simulations, we do not observe any pseudosynchronization of strongly tidally heated exoplanets with magma ocean. Since the solidus temperature increases with pressure, the melt is emerging only under the lithosphere and---in our model---it remains in the same place as long as the local temperature exceeds solidus. The zone of substantially reduced viscosity and rigidity is thus concentrated only in the upper mantle, while the lower mantle remains solid and maintains relatively high average viscosity. As the molten layer is almost nondissipative, its formation effectively reduces the volume in which the mechanical energy transforms to heat. This mechanism is then responsible for the reduced rate of tidal dissipation and orbital evolution \citep[see also][]{henning14}.

4) The long-term thermal-orbital evolution of tidally loaded rocky exoplanets is strongly interconnected with the evolution of their spin rate. For higher than synchronous spin-orbit resonances, the thermal state promptly evolves into an equilibrium, which is stable as long as the planet remains in the same resonance. The equilibrium temperature profile ensures that the heat sources are effectively compensated by the heat loss. Since the tidal dissipation at higher than synchronous resonances depends mainly on the spin rate and only weakly on the eccentricity, the thermal equilibria are stable for a considerable time ($\sim\unit{Gyr}$ in some cases), almost independently of the evolving orbital parameters. Each transition between spin-orbit resonances is then accompanied by a transition between thermal equilibria. Once the planet despins to synchronous rotation, the tidal heat rate becomes sensitive to the orbital eccentricity and the planetary mantle cools down gradually.\\

Understanding the complex relation between the interior dynamics and the quantities which can be theoretically measured may help us to better constrain the conditions on extrasolar worlds. Although the spin rate of the studied low-mass exoplanets is currently beyond the limits of observational techniques, it might be measurable by the upcoming ground-based or space-based missions, such as JWST or E-ELT \citep[see, e.g.,][for the discussion of observational prospects of Proxima Cetauri b]{kane17b}. The surface tidal heat flux, which is not directly measurable, may affect the rate of volcanic activity and outgassing. The footprints of increased interior heating would then be observable in the planet's transmission spectra or, possibly, in the infrared lightcurves \citep[e.g.,][]{demory12,selsis13,meadows18}. Observational constraints on the spin rate and on the heat production, together with precise assessment of the orbital eccentricity, are crucial in determining the geophysical properties of tidally loaded exoplanets.

\section{Acknowledgements}
We would like to thank Dr. Michael Efroimsky for his valuable comments on the manuscript. The research leading to these results received funding from the Czech Science Foundation through project No.~19-10809S (all authors) and from the Charles University through project SVV 115-09/260581 (M.W.).

\appendix

\section{Outline of the normal mode theory} \label{app1}

Assuming an incompressible, elastic and layered spherical planet subjected to external body force, the incremental deformation can be expressed as a sum of radial and lateral terms,

\footnotesize
\begin{equation}
\mathbf{u} = \sum_{n=0}^{\infty} \left[U_n(r)\,\mathcal{P}_n(\cos\theta)\,\mathbf{e}_r + V_n(r)\, \frac{\partial}{\partial\theta} \mathcal{P}_n(\cos\theta)\, \mathbf{e}_{\theta}\right],
\end{equation}\normalsize\\
the additional potential induced by the deformation can be decomposed into

\footnotesize
\begin{equation}
\delta\phi = \sum_{n=0}^{\infty} \Phi_n(r)\, \mathcal{P}_n(\cos\theta)
\end{equation}\normalsize\\
and the incremental pressure is

\footnotesize
\begin{equation}
\delta p = \sum_{n=0}^{\infty} \Pi_n(r)\, \mathcal{P}_n(\cos\theta)\; .
\end{equation}\normalsize\\
In the above expressions, $\mathcal{P}_n(\cos\theta)$ are Legendre polynomials and $\mathbf{e}_r$, $\mathbf{e}_{\theta}$ are unit vectors in the radial and lateral (eastward) directions, respectively. As a consequence of the spherical harmonic decomposition, the set of governing equations \citep[e.g.,][]{sabadini04}, which ensures the conservation of mass and momentum in the continuum, as well as the constitutive equation for an elastic material, can be rewritten into a set of ordinary differential equations of the form

\footnotesize
\begin{equation}
\dot{\mathbf{Y}} = \mathbb{A}\mathbf{Y}, \label{eq:odeset}
\end{equation}\normalsize\\
where

\footnotesize
\begin{equation}
\mathbf{Y} = \Big(U_n,V_n,T_{rn},T_{\theta n},\Phi_n,Q_n\Big)^t \label{eq:solvec}
\end{equation}\normalsize\\
with $U_n$, $V_n$ and $\Phi_n$ introduced above and the other variables defined as

\footnotesize
\begin{align}
T_{rn}(r) &= -\Pi_n + 2\mu \dot{U}_n \; , \label{eq:trn}\\[0.5ex]
T_{\theta n}(r) &= \mu \left(\dot{V}_n-\frac{1}{r}V_n + \frac{1}{r}U_n \right) \; , \label{eq:ttn}\\[0.5ex]
Q_n(r) &= \dot{\Phi}_n + \frac{n+1}{r} \Phi_n + 4\pi\mathcal{G}\rho U_n \; .
\end{align}\normalsize\\
Here, $\rho$ is the mean density at radius $r$ and symbol $\mu$ stands for static rigidity. The set of governing equations is constrained by boundary conditions prescribed at each interior interface, at the surface, and in the centre. Specifically, in the case of tidal loading, the boundary conditions at the surface are \citep{takeuchi62,sabadini04}

\footnotesize
\begin{equation}
\begin{aligned}
T_{rn}(R) &= 0\; , \\[0.5ex]
T_{\theta n}(R) &= 0\; , \\[0.5ex]
Q_n(R) &= -\frac{2n+1}{R}\; .
\end{aligned}
\end{equation}\normalsize\\
In the centre, we only require regularity of the solution. The interior boundary conditions depend on the type of interfaces between the layers. Here, we consider each transition between layers as a solid-solid interface, which yields continuity of all but one components of the vector (\ref{eq:solvec}) and a step in $Q_n$ due to boundary deflections, i.e.,

\footnotesize
\begin{equation}
\begin{gathered}
\big[U_n\big]^+_- = \big[V_n\big]^+_- = \big[T_{rn}\big]^+_- = \big[T_{\theta n}\big]^+_- = \big[\Phi_n\big]^+_- = 0 \\[0.5ex]
\big[Q_n\big]^+_- = -\frac{n+1}{r} \Phi_n \\[0.5ex]
\end{gathered}
\end{equation}\normalsize\\
When applying the correspondence principle and shifting from the elastic to the viscoelastic problem, this choice of boundary conditions also means that each liquid layer needs to be considered as viscous. The incorporation of inviscid layers is, however, straightforward \citep{wu82}.\\

The general solution to the set of ordinary differential equations (\ref{eq:odeset}) in layer $j\in[1,N]$ is a superposition of six linearly independent solutions,

\footnotesize
\begin{equation}
\mathbf{Y}(r) = \sum_{i=1}^{6} C_i^{(j)} \mathbf{y}_i(r)\; , \label{eq:ydef}
\end{equation}\normalsize\\
where $C_i^{(j)}$ are layer-dependent constants given by the boundary conditions. Both the constants $C_i^{(j)}$ and the individual solutions $\mathbf{y}_i(r)$ of the viscoelastic problem attain complex values, which contain the information on the amplitude of deformations, stresses and potential, alike with the physical lagging caused by attenuation in the medium. While the phase lag between the strain and stress enables us to enumerate the energy dissipation anywhere inside the planet, the lagging between the external and additional gravitational potential presents a key parameter entering the spin and orbital evolution equations (\ref{eq:kaula_a}), (\ref{eq:kaula_e}) and (\ref{eq:rot}). Specifically, the complex tidal Love number $\bar{k}_l(\omega)$ is related to the constants $C_i^{(j)}$ by

\footnotesize
\begin{equation}
\bar{k}_n(\omega) = -C_3^{(N)}R^n - C_6^{(N)}R^{-n-1} - 1 \; .
\end{equation}\normalsize

\section{Parameter dependence of tidal torque} \label{app2}

Degree $2$ secular tidal torque acting on a planet with zero obliquity can be expressed from equation (\ref{eq:rot}) as an infinite sum

\footnotesize
\begin{equation}
\mathcal{T} = \sum_{q=-\infty}^{\infty} \mathcal{T}_{220q} = -K\; \sum_{q=-\infty}^{\infty} \Big[\mathcal{G}_{20q}\Big]^2 \rm{Im}\Big\{\bar{k}_{220q}(\omega_{220q})\Big\}\; , \label{eq:apptorque}
\end{equation}\normalsize
with
\footnotesize
\begin{equation*}
    K = \frac32 \frac{\mathcal{G}m_*^2 R^5}{a^6}\; .
\end{equation*}\normalsize\\
In a simplified case of homogeneous interior governed by the Maxwell rheology, we may write the imaginary part of the complex Love number as \citep[e.g.,][]{castillo11}

\footnotesize
\begin{equation}
\rm{Im}\Big\{\bar{k}(\omega)\Big\} = -\frac{57}{4} \left(\frac{\rho g R \eta \omega}{\mu^2} + \frac{19\eta\omega}{\mu} + \frac{361}{4}\frac{\eta\omega}{\rho g R} + \frac{\rho g R}{\eta\omega}\right)^{-1}\; , \label{eq:maxtorque}
\end{equation}\normalsize\\
where $\omega$ is the tidal frequency of the given mode (equation (\ref{eg:freqs})).\\

To emphasize the frequency dependence of this expression and analyze the stability of higher spin-orbit resonances, we further simplify the notation of equation (\ref{eq:maxtorque}) and rewrite it to the form

\footnotesize
\begin{equation}
    \rm{Im}\Big\{\bar{k}(\omega)\Big\} = -K'\, f(\omega) = -\frac{K'}{A\omega+\frac{1}{\omega}} \; , \label{eq:kink}
\end{equation}\normalsize
where we substituted
\footnotesize
\begin{equation*}
    K' = \frac{57}{4} \frac{\eta}{\rho g R} \qquad \mbox{and} \qquad A = \frac{\eta^2}{\mu^2}\left(1+\frac{19}{2}\frac{\mu}{\rho g R}\right)^2
\end{equation*}\normalsize\\
Function $f(\omega)$ is responsible for a kink-shaped torque around spin-orbit resonances, which ensures their stability in viscoelastic rheological models \citep[][]{makarov13,noyelles14}. Specifically, the functional dependence of $f(\omega)$ on the frequency $\omega$ is given by the coefficient $A$, i.e., by a combination of rheological and physical parameters of the planet. Depending on the relative magnitude of these quantities, we may delimit two regions in the parameter space:

\footnotesize
\begin{align*}
    &\frac{19}{2}\frac{\mu}{\rho g R} \ll 1 &\mbox{self-gravity-dominated regime} \qquad &A\approx\tau_{\rm{M}}^2 \\[0.5ex]
    &\frac{19}{2}\frac{\mu}{\rho g R} \gg 1 &\mbox{rheology-dominated regime} \qquad &A\approx\left(\frac{19}{2}\frac{\eta}{\rho g R}\right)^2
\end{align*}\normalsize\\
For Earth-like planets, the boundary between the self-gravity-dominated and rheology-dominated regions lies at $\mu\sim\unit[10^{10}]{Pa}$. In the rheology-dominated regime, tidal torque is determined solely by the planet's viscosity and does not depend on the rigidity. The same behaviour is predicted also for small bodies, such as asteroids \citep{efroimsky15}, as long as their reaction can be described by the Maxwell rheology. In the self-gravity-dominated regime, on the other hand, the shape of the kink function (\ref{eq:kink}) is determined by the Maxwell time $\tau_{\rm{M}}=\frac{\eta}{\mu}$. While the parameter dependence of $A$ in the two regimes might be different, the same value of $A$ should always yield the same behaviour of $f(\omega)$ around zero.

Depending on the magnitude of $A$ and on the loading frequency $\omega$, we may also distinguish two limit cases,

\footnotesize
\begin{align*}
    &A\omega^2\ll1 \qquad \Rightarrow \qquad f(\omega)\approx\omega\; , \\[0.5ex]
    &A\omega^2\gg1 \qquad \Rightarrow \qquad f(\omega)\approx\frac{1}{A\omega}\; .
\end{align*}\normalsize\\
The first case, if attained in the self-gravity-dominated regime, corresponds to the weak friction approximation, used originally for the description of tides in binary stars \citep[e.g.,][]{alexander73,hut81}, or to the constant time lag model \citep[e.g.,][]{mignard79,correia10}.\\

The secular tidal torque (\ref{eq:apptorque}) is, in fact, a weighted sum of the kink-shaped functions $f(\omega)$, centered around frequencies $\omega_{220q}$. Standing alone, each kink would cross zero exactly at a half-integer spin-orbit resonance, such as 1:1, 3:2, and so on. However, when multiplied by the coefficients $\left[\mathcal{G}_{20q}(e)\right]^2$ and summed together, the positions of their zero crossings get slightly shifted from the exact resonances --- and, in some cases, they do not cross zero at all. Our parametric studies in Sections \ref{sec:results1} and \ref{sec:results2} show the highest spin-orbit resonance for which the tidal torque still crosses zero.

For the sake of illustration, let us derive a stability criterion for the 3:2 resonance. Assuming a homogeneous spherical planet governed by the Maxwell rheology, we may rewrite the secular tidal torque expanded to the second order in orbital eccentricity as

\footnotesize
\begin{equation}
    \mathcal{T}_2 = K_2 \left\{f(2n-2\dot{\theta}) + e^2 \left[\frac14f(n-2\dot{\theta})-5f(2n-2\dot{\theta})+\frac{49}{4}f(3n-2\dot{\theta})\right]\right\}\; ,
\end{equation}\normalsize
where
\footnotesize
\begin{equation}
    K_2 = K\, K'\; .
\end{equation}\normalsize\\
Frequencies $(n-2\dot{\theta})$, $(2n-2\dot{\theta})$ and $(3n-2\dot{\theta})$ correspond to the 1:2, 1:1 and 3:2 resonances, respectively. Note that in the low-eccentricity case, when $e\lesssim0.1$, the largest term in the expansion is the 1:1 resonance with prefactor $(1-5e^2)$. To the right from this resonance, $f(2n-2\dot{\theta})$ is negative. Since the stability of the 3:2 resonance requires that $\mathcal{T}_2$ crosses zero in its vicinity, we are seeking the parameters for which the maximum of the 3:2 kink is non-negative. If we neglect the contribution of the term corresponding to the 1:2 resonance, the problem reduces to a comparison between the 1:1 and the 3:2 components. The maximum of the 3:2 component lies at $\dot{\theta} = \frac32 n - \frac12\sqrt{\frac{1}{A}}$. Thus, the 3:2 resonance is theoretically stable, whenever

\footnotesize
\begin{equation}
    (1-5e^2)\, f\left(-n+\sqrt{\frac{1}{A}}\right) + \frac{49}{4} e^2\, f\left(\sqrt{\frac{1}{A}}\right) > 0\; . \label{eq:ineq}
\end{equation}\normalsize\\
If we further require that the maximum of the 3:2 component is positioned to the right from the 1:1 resonance, the above inequality is solved with

\footnotesize
\begin{equation}
    \sqrt{A} > \frac{29e^2+4+4\sqrt{1-10e^2-({2001}/{16})e^4}}{49 e^2 n} \approx \frac{8}{49e^2n}\; . \label{eq:stable32}
\end{equation}\normalsize\\
We recall that this condition has been derived with the explicit assumption of small eccentricity ($e\lesssim0.1$) and holds only for the Maxwell rheology. A similar analysis can be also performed for higher eccentricities and other spin-orbit resonances (with a higher-order expansion of the secular tidal torque), as well as for different rheological models. Specifically, the additional terms in the Andrade model, when compared to the Maxwell model, make the stability criterion stricter and more complex than in this illustrative case.

\bibliography{litera}{}
\end{document}